\documentclass[twocolumn]{aastex631}

\usepackage{comment}
\usepackage{quoting}
\usepackage{amsmath}
\usepackage{enumitem}
\usepackage{booktabs}
\usepackage{subfigure}
\usepackage{multirow, booktabs}

\shortauthors{Maben et al.}
\shorttitle{Asteroseismology of CDGs}

\begin{document}

\title{Asteroseismology of Carbon-Deficient Red Giants: Merger Products of Hierarchical Triple Systems?}

\author[0000-0001-9974-1754]{Sunayana Maben}
\affil{CAS Key Laboratory of Optical Astronomy, National Astronomical Observatories, Chinese Academy of Sciences, Beijing 100101, Peoples Republic of China}

\author{Simon W. Campbell}
\affil{School of Physics and Astronomy, Monash University, Clayton, Victoria, Australia}
\affil{ARC Centre of Excellence for Astrophysics in Three Dimensions (ASTRO-3D), Australia}

\author[0000-0001-5222-4661]{Timothy R. Bedding}
\affiliation{Sydney Institute for Astronomy, School of Physics, University of Sydney NSW 2006, Australia}

\author[0000-0002-8980-945X]{Gang Zhao}
\affil{CAS Key Laboratory of Optical Astronomy, National Astronomical Observatories, Chinese Academy of Sciences, Beijing 100101, Peoples Republic of China}
\affil{School of Astronomy and Space Science, University of Chinese Academy of Sciences, Beijing 100049, Peoples Republic of China}

\author[0000-0003-0929-6541]{Madeline Howell}
\affil{Department of Astronomy, The Ohio State University, 140 W 18th Avenue, Columbus, OH 43210, USA}

\author{Yerra Bharat Kumar}
\affil{Indian Institute of Astrophysics, 100ft road Koramangala, Bangalore, 560034, India}

\author[0000-0001-9246-9743]{Bacham E. Reddy}
\affil{Department of Physics, Indian Institute of Technology Jammu, Jammu 181221, India}

\correspondingauthor{Sunayana Maben and Gang Zhao}
\email{smaben@nao.cas.cn; gzhao@nao.cas.cn}
\begin{abstract}
Carbon-deficient giants (CDGs) are a rare and chemically peculiar class of stars whose origins remain under active investigation. We present an asteroseismic analysis of the entire known CDG population, selecting 129 stars observed by $Kepler$, K2, and TESS to obtain seismic constraints. We detect solar-like oscillations in 43 CDGs. By measuring $\nu_{\rm max}$ and applying seismic scaling relations, we determine precise masses for these stars, finding that 79\% are low-mass ($M \lesssim 2~M_\odot$). The luminosity distribution is bimodal, and the CDGs separate into three chemically and evolutionarily distinct groups, characterized by clear trends in sodium and CNO abundances, $\alpha$-element enhancement, and kinematics. We find that two of these groups are only distinguished by their initial $\alpha$-element abundances, thus effectively reducing the number of groups to two. Lithium enrichment is common across all groups, linking CDGs to lithium-rich giants and suggesting a shared evolutionary origin. We find that spectroscopic $\log g$ is systematically offset from seismic values. Group~1 CDG patterns are most consistent with formation through core He-flash mixing, while the more massive and more chemically processed Groups~2 and 2$\alpha$ likely formed through mergers involving helium white dwarfs, possibly in hierarchical triples. Pollution from AGB stars appears very unlikely, given the unchanged [C+N+O] abundance across all groups.

\end{abstract}

\section{Introduction} \label{sec:introduction}

The study of carbon-deficient giants (CDGs) has a long history. More than a century ago, \cite{Cannon1912} noticed that the red giant star HR~885 had an unusual spectrum, where the G-band—caused by CH molecular absorption—was not well defined, unlike in other G- and K-type stars. \cite{Bidelman1951} closely examined the spectrum of this star and concluded that the G-band was absent. Later, \cite{Bidelman1973} identified a group of G and K giants with extremely low carbon abundances, much lower than expected from the first dredge-up process. These stars, known as weak G-band (wGb) stars, exhibit very weak or absent CH molecular absorption at 4300~\AA. More recently, \cite{Bond2019} added five more CDGs to the previously known list of 39 wGb stars, bringing the total to 44. High-resolution spectroscopic studies (e.g., \citealt{Adamczak2013}; \citealt{Palacios2016}) confirmed that these stars not only have very low carbon abundances but also show low carbon isotopic ratios ($\rm^{12}C/^{13}C \approx 3\text{--}4$) and enhanced nitrogen. Some of them are also enriched in lithium or sodium.

Earlier studies suggested that wGb stars are intermediate-mass stars ($M = 2.5\text{--}5.0~\rm M_{\odot}$) based on their positions in the Hertzsprung–Russell diagram (HRD). Most were found to be in the subgiant branch (SGB) or red giant branch (RGB) phases, and a few in the red clump (RC) phase (e.g., \citealt{Palacios2012}; \citealt{Bond2019}).
However, new findings have revised this understanding. \cite{Maben2023a} and \cite{Maben2023b} expanded the known sample of CDGs by a factor of three by identifying around 100 additional stars using APOGEE survey data \citep{Abdurrouf2022}. Their analysis of 15 CDGs in the $Kepler$ field, incorporating asteroseismic mass determinations, revealed that these stars are predominantly low-mass ($M \lesssim 2$~M$_{\odot}$), in contrast to earlier findings. Remarkably, they also found that these CDGs are almost exclusively in the RC phase. Furthermore, a distinct pattern in their luminosity distribution led to the classification of CDGs into three groups. Possible formation scenarios for these stars include mergers between helium white dwarfs (HeWDs) and RGB stars, as well as binary mass transfer from asymptotic giant branch (AGB) stars. The overlap between CDGs and lithium-rich giants suggests that they may share similar evolutionary origins \citep{Maben2023b}.

In this work, we extend the asteroseismic analysis of \citet{Maben2023b} to the full sample of 158 known CDGs \citep{Bidelman1951, Bidelman1973, Bond2019, Holanda2023, Maben2023a, Maben2023b}, utilizing data from the $Kepler$, K2 and TESS missions. By combining asteroseismic information with astrometric, photometric, and spectroscopic data, we aim to refine mass estimates and to establish a more comprehensive evolutionary framework for CDGs. This approach is crucial because conventional spectroscopic analyses tend to underestimate $\log g$ in CDGs \citep{Palacios2016}, directly affecting mass determinations and highlighting the importance of asteroseismic constraints for both reliable mass estimates and accurate surface gravity measurements. Given the recent paradigm shift in our understanding of these stars, this study systematically classifies CDGs and investigates their formation mechanisms. By using multiple constraints on a larger sample, we provide new insights into their nature, formation, and evolution.

\section{Sample selection} \label{sec:sample-selection}

In the current study, we focus on the TESS, $Kepler$ and K2 fields because we are interested in obtaining asteroseismic constraints. We cross-matched the entire list of CDGs with stars observed by those three missions, resulting in a sample of 129 stars. Among these, 128 CDGs have TESS light curves, 17 have $Kepler$ light curves, and 4 have K2 light curves.

We searched the literature for stars with previously published global asteroseismic parameters ($\nu_{\text{max}}$ and $\Delta\nu$) and identified 27 stars. Among these, 16 are located in the $Kepler$ field, with 15 included in the \citet{Maben2023b} study. All 16 $Kepler$ stars have spectroscopic data from APOGEE and asteroseismic parameters derived from the $Kepler$ light curves. Specifically, $\nu_{\text{max}}$ and $\Delta\nu$ measurements are available for all 16 stars \citep{Mosser2014,Vrard2016,Yu2018,Yu2020}, while $\Delta\Pi_1$ values are reported for 11 of them \citep{Mosser2014,Vrard2016}.
Among the remaining 11 stars with published asteroseismic parameters, all have $\nu_{\text{max}}$ derived from TESS light curves \citep{Hon2021, Zhou2024}, and 3 also have $\Delta\nu$ reported in \citet{Zhou2024}.

In summary, we have 27 CDGs with global asteroseismic parameters determined in the literature, and 102 CDGs that do not. We aim to determine asteroseismic parameters for as many of these 129 CDGs as possible.

\section{Methodology} \label{sec:methodology}

Asteroseismology typically uses both the large frequency separation ($\Delta\nu$) and the frequency of maximum oscillation power ($\nu_{\text{max}}$) to determine stellar properties such as mass and radius \citep[e.g.,][]{Chaplin2013}. However, in this study, we focus solely on measuring $\nu_{\text{max}}$ because the determination of $\Delta\nu$ is highly uncertain due to the limited amount of observational data available for most stars in our sample. This limitation is especially significant for the 111 CDGs with TESS data alone, where 79 stars (71$\%$) have been observed in fewer than five TESS sectors, limiting the signal-to-noise ratio (SNR) and frequency resolution of the power spectra.

Similar challenges have been noted in previous studies, where data limitations restrict reliable $\Delta\nu$ estimation \citep[e.g.,][]{Hon2021, Stello2022, Howell2025}. Given that $\nu_{\text{max}}$, when combined with temperature and luminosity, has been shown to be sufficient for deriving accurate asteroseismic masses \citep[e.g.,][]{Hon2021, Howell2022, Malla2024}, we adopt this approach in our analysis.

The steps to measure $\nu_{\text{max}}$ can be summarized as follows:
\begin{enumerate}[label=(\roman*),itemsep=1pt,topsep=1pt]
    \item Calculate an initial estimate of $\nu_{\text{max}}$ based on the standard scaling relation (Section~\ref{sec:methodology-initial-numax})
    \item Calculate the power density spectrum of the light curve (Section~\ref{sec:methodology-calculating-PDS})
    \item Measure  $\nu_{\text{max}}$ and its associated uncertainty (Section~\ref{sec:methodology-measuring-numax})
\end{enumerate}

\subsection[Initial estimate of numax]{Initial estimate of $\nu_{\text{max}}$}  \label{sec:methodology-initial-numax}
An initial estimate of $\nu_{\text{max}}$ is required for each star, as steps (ii)$-$(iii) depend on this parameter. To obtain this estimate, we employed the standard scaling relation \citep{Brown1991, Kjeldsen1995}:
\begin{equation}
\nu_{\rm max} \simeq \nu_{\rm max,\odot} \left( \frac{g}{g_{\odot}} \right) \left( \frac{T_{\rm eff}}{T_{\rm eff,\odot}} \right)^{-1/2},
\label{eq:initial_numax}
\end{equation}
where we adopted $\nu_{\rm max,\odot} = 3090 \pm 30~\mu$Hz \citep{Huber2011}, $T_{\rm eff,\odot} = 5772$~K \citep{Andrej2016}, and $\log g_{\odot} =4.44$ \citep{Morel2014} as our solar reference values.

The effective temperatures ($T_{\rm eff}$) and surface gravities ($g$) for the CDGs were taken from the literature \citep{Adamczak2013, Palacios2016, Holanda2023, Maben2023a, Maben2023b, Holanda2024}. In cases where these parameters were not available, we used values from $Gaia$~DR3 \citep{Gaia2016, Gaia2023}. The available values of $\log g$  are not particularly accurate---indeed, refining these values using asteroseismology is one our main aims---but they suffice for giving an initial estimate of $\nu_{\rm max}$.

\subsection{Pre-processing light curves and calculating power spectra} \label{sec:methodology-calculating-PDS}

We downloaded all available light curves for the CDGs from the Mikulski Archive for Space Telescopes\footnote{\url{https://mast.stsci.edu/portal/Mashup/Clients/Mast/Portal.html}} (MAST) using the Python package \texttt{lightkurve} \citep{Lightkurve2018}. These light curves have been extracted and detrended using mission-specific pipelines: $Kepler$ \citep{Jenkins2010}, K2 \citep{Vanderburg2014, Luger2016}, and TESS \citep{Jenkins2016, Caldwell2020, Huang2020a, Huang2020b}. We included all TESS data available up to the end of Sector~81 (10 August 2024). We performed a quality assessment of each light curve through visual inspection, evaluating them for systematic trends, instrumental artifacts, and noise levels. We then selected those light curves with minimal contamination for further analysis, ensuring that only high-quality data from $Kepler$ quarters, K2 campaigns, and TESS sectors were used. This resulted in a modest fraction of the data being discarded (161 sectors for 40 different stars).

To eliminate low-frequency variations caused by stellar activity and instrumental noise—often not corrected for by PDCSAP from the $Kepler$ and TESS missions \citep{Smith2012, Jenkins2016}—we applied a high-pass filter to each light curve, as follows. We divided $Kepler$ light curves into two segments per quarter and we divided TESS light curves into four half-orbits per sector (giving segments of approximately 7 days each), to account for flux jumps caused by momentum dumping. The high-pass filter was applied to each segment by dividing the original flux by a smoothed version of the light curve, obtained through convolution with a Gaussian kernel. The filter cut-off, set by the width of the Gaussian, was varied from star to star, based on the estimated $\nu_{\text{max}}$ from the previous step. The cut-off frequency was set well below this value to ensure that the filter did not interfere with the oscillations under study. This adaptive filtering technique allowed us to effectively reduce low-frequency noise while preserving higher frequency oscillations.

Next, we concatenated the light curves from all available sectors and calculated power spectra up to the Nyquist frequency. The power (in ppm\(^2\)) was converted to power density (in ppm\(^2 \, \mu\text{Hz}^{-1}\)) by multiplying by the effective time span of the observations \citep{Kjeldsen1995}.

\begin{figure}[]
        \centering
        \includegraphics[width=1.\linewidth]{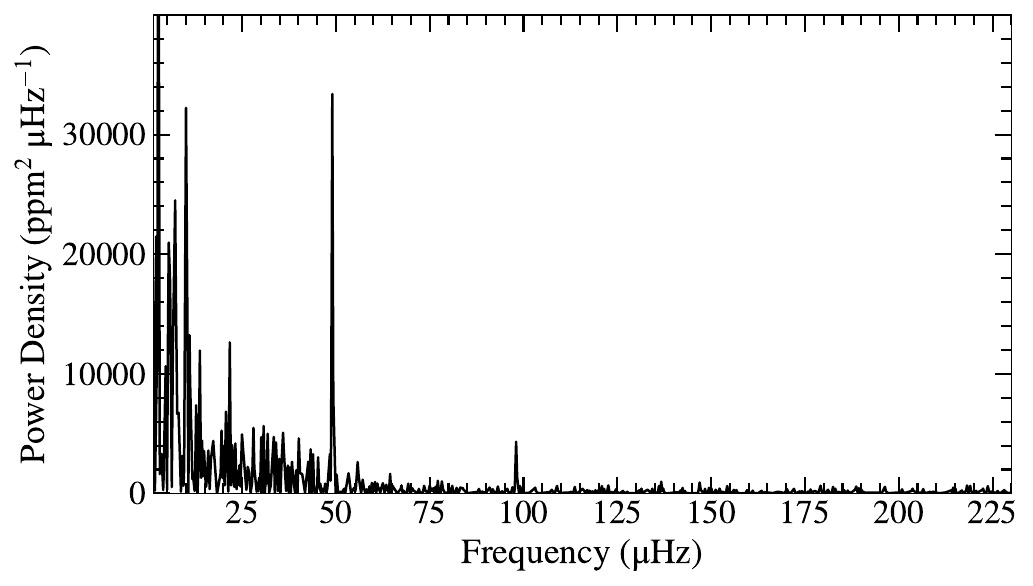}
        \caption{Power spectrum of HD~124721, a CDG with an anomalous peak at 50~$\mu$Hz, calculated from a single sector of TESS data. The solar-like oscillations are centered at 35~$\mu$Hz.}
        \label{fig:HD124721-anomalous-peaks}
    \end{figure}

We identified three stars, HD~124721, HD~40402, and BD+5~593, that exhibited anomalous peaks in their power spectra, likely indicative of contamination by a close binary \citep{Colman2017}. Notably, HD~124721 and HD~40402 have been found to be in either multiple systems or binary systems in different catalogues \citep{Mason2001, Dommanget2002, Badry2021}. For example, Figure~\ref{fig:HD124721-anomalous-peaks} shows the power spectrum of HD~124721, calculated from a single sector of TESS data, which features an anomalous high-amplitude peak and its second harmonic. These strong peaks were removed to refine our analysis. Such signals are typically attributed to either chance alignments with unrelated binary systems or physical associations in hierarchical triples \citep{Colman2017}. This observation highlights the complexity of stellar interactions and the necessity for careful interpretation of power spectra in the context of binary and multiple star systems.

\subsection[Measuring numax]{Measuring $\nu_{\text{max}}$} \label{sec:methodology-measuring-numax}

To measure $\nu_{\text{max}}$--and its associated uncertainty--we used the \texttt{pyMON} pipeline\footnote{\url{https://github.com/maddyhowell/pyMON}} \citep{Howell2025}, which is directly based on \texttt{pySYD} \citep{Huber2009, Chontos2022}. The \texttt{pyMON} pipeline is ideal for our purpose because it is optimized for measuring $\nu_{\text{max}}$ in low SNR data, and for low-$\nu_{\text{max}}$ giants \citep{Howell2025}.

\begin{figure}[]
        \centering
        \includegraphics[width=\linewidth]{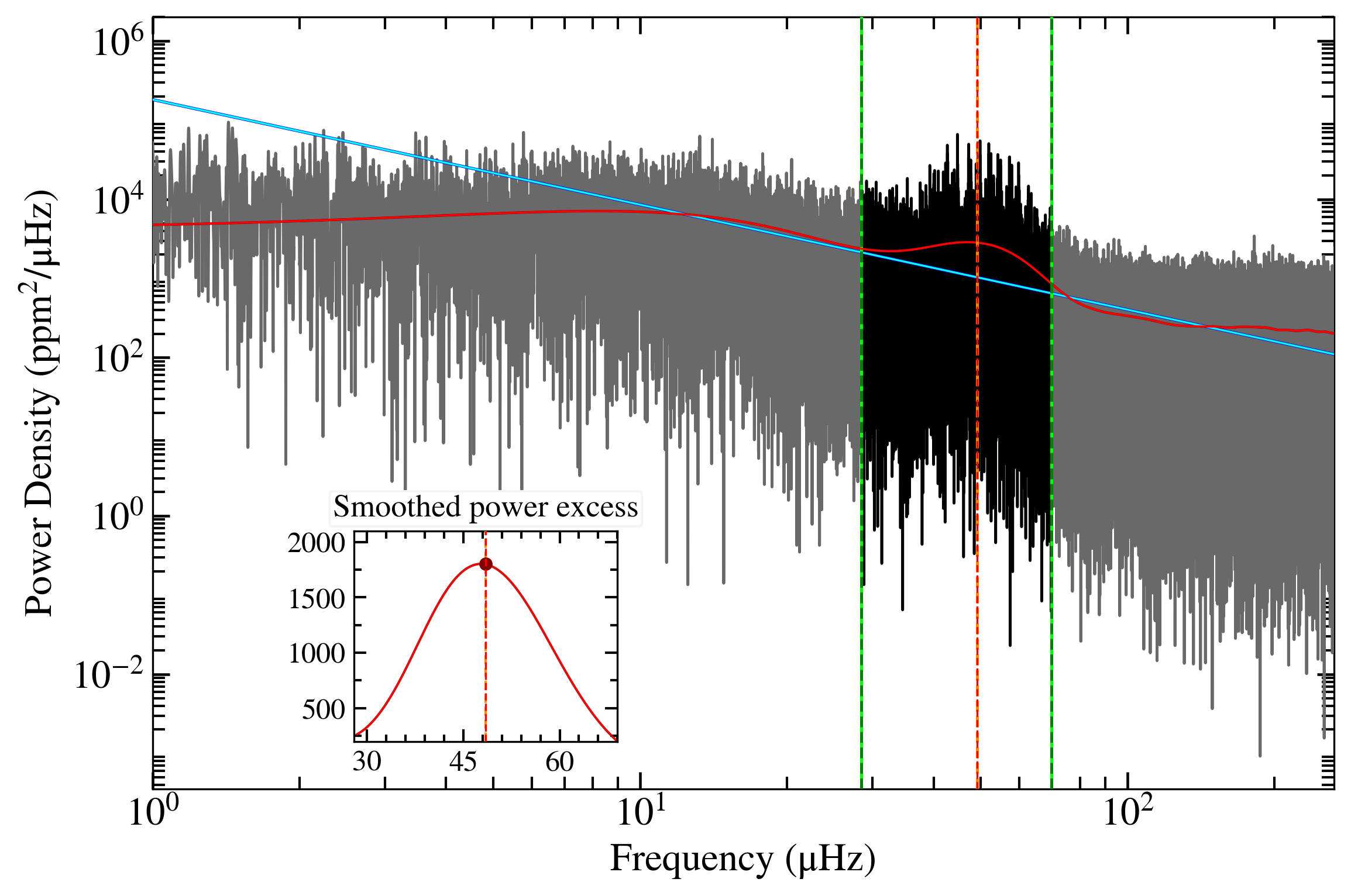}
        \caption{Power spectrum of 2M19125144+3850261 (grey), including the smoothed power spectrum (red) and the linear background fit (blue). The power excess (black) is positioned between two vertical green lines, which mark the window. The vertical red dashed line represents the measured $\nu_{\rm max}$ at 49~$\mu$Hz. The \texttt{pyMON} pipeline estimates $\nu_{\rm max}$ as the frequency of maximum power in the background-corrected smoothed power spectrum (shown in inset).}
        \label{fig:KIC-3736289-pyMON-pipeline}
    \end{figure}

The initial estimate of $\nu_{\text{max}}$ is used by \texttt{pyMON} to define the power excess window where oscillations are expected, which can be further refined with optional lower and upper frequency limits if the data are noisy. The background is removed from the power spectrum by subtracting a linear model derived from the intersection between the power excess window boundaries (see Figure~\ref{fig:KIC-3736289-pyMON-pipeline} for an example). This tends to work more consistently for stars with low $\nu_{\text{max}}$, where there are too few low-frequency data points to reliably constrain a Harvey-like granulation model  \citep{Howell2025}. Following this, the spectrum is heavily smoothed guided by a $\Delta\nu$ estimate based on the $\Delta\nu$--$\nu_{\text{max}}$ scaling relation, as described by \cite{Stello2009}. The frequency corresponding to the maximum power within the power excess window is adopted as $\nu_{\text{max}}$. The pipeline performs this analysis in a single pass, relying on the initial $\nu_{\text{max}}$ estimate, consistent with its \texttt{pySYD}-based approach. To determine the uncertainty, stochastic noise is introduced into the power spectrum, and $\nu_{\text{max}}$ is re-estimated. This process is repeated 500 times, and the standard deviation of the resulting $\nu_{\text{max}}$ distribution serves as the uncertainty estimate, consistent with the approach of \cite{Huber2009}.

Our analysis of 129 CDGs using TESS, $Kepler$, and K2 data resulted in clear detections of solar-type oscillations in 43 CDGs, which constitutes $33$\% of the sample. The detailed detection statistics are presented in Section~\ref{sec:numax-measurements}.

With $\nu_{\text{max}}$ determined, we applied the following asteroseismic scaling relation to derive stellar mass \citep{Stello2008}:
\begin{equation}
\left( \frac{M}{M_\odot}\right) \simeq \left(\frac{\nu_{\rm max}}{\nu_{\rm max,\odot}}\right) \left(\frac{L}{L_{\odot}}\right) \left(\frac{T_{\rm eff}}{T_{\rm eff,\odot}}\right)^{-7/2}. \label{eq:mass3}
\end{equation}
Note that this uses the luminosity determined as described in Section~\ref{sec:luminosities} and the spectroscopic effective temperature, but not the spectroscopic estimate of surface gravity.

\section{Results and Discussion}\label{sec:results}
\subsection[Comparison with published numax values]{Comparison with published $\nu_{\rm max}$ values} \label{sec:comparision-with-published-numax}

\begin{figure}
        \centering
        \includegraphics[scale=5,width=\linewidth]{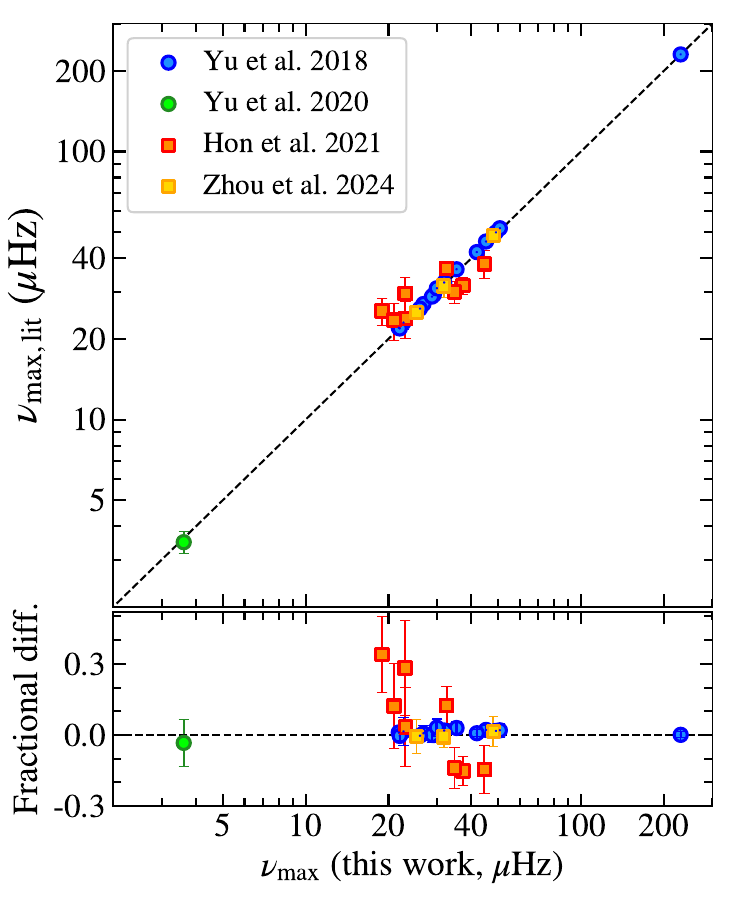}
         \caption{Comparison of the $\nu_{\text{max}}$ values obtained using \texttt{pyMON} (this work) with literature values ($\nu_{\text{max, lit}}$) from $Kepler$ \citep{Yu2018,Yu2020} and TESS \citep{Hon2021,Zhou2024} (see Table~\ref{tab:validation}). The diagonal black line indicates the one-to-one relation. The bottom panel shows the fractional difference, defined as $(\nu_{\text{max, lit}} - \nu_{\text{max}}) / \nu_{\text{max}}$. Apart from the \citet{Hon2021} sample, the agreement between our values and those in the literature is excellent (see text for details).}
        \label{fig:pyMON-numax-comparision}
    \end{figure}

As a check, we present a comparative analysis for 27 CDGs between the published $\nu_{\text{max}}$ values and our values obtained using the \texttt{pyMON} pipeline (see Table~\ref{tab:validation} and Figure~\ref{fig:pyMON-numax-comparision}). Our goal is to assess how our measured values align with those reported in the literature.

When compared with \citet{Yu2018,Yu2020}, who used the four-year $Kepler$ data, and \cite{Zhou2024}, who used 2-minute cadence TESS data from sectors 1--60, the mean fractional difference is $0.7\%$ with a standard deviation of $1.5\%$. This close agreement demonstrates that our $\nu_{\text{max}}$ measurements are consistent with those derived from the same or similar datasets.

In contrast, the comparison with \citet{Hon2021}, who employed 30-minute cadence TESS light curves from sectors 1--26, shows larger discrepancies. These arise primarily from the more limited data coverage available in that study, which affects the precision of the $\nu_{\text{max}}$ measurements and leads to increased scatter. Given these differences in data quality and coverage, we exclude the \citet{Hon2021} sample from our main quantitative comparison, but note that our results remain broadly consistent within the uncertainties.

For stars observed with $Kepler$, and with TESS at 2-minute cadence, the typical relative uncertainties in the $\nu_{\text{max}}$ measurements range from 1--5\%. For most stars, the differences between our values and those from the literature fall within 1--2$\sigma$ of the combined uncertainties, indicating that the observed scatter is consistent with expected measurement errors.

These $\nu_{\text{max}}$ differences propagate directly into the seismic mass estimates, as evidenced by our comparative analysis of masses derived using both \texttt{pyMON} and literature $\nu_{\text{max}}$ values (see Section~\ref{sec:masses}). According to the standard asteroseismic scaling relation (Equation~(\ref{eq:mass3})), the seismic mass scales linearly with $\nu_{\text{max}}$, meaning that a proportional offset in $\nu_{\text{max}}$ would translate to the same percentage difference in the derived masses, assuming other parameters remain constant.

In summary, our measured $\nu_{\text{max}}$ values, derived using the \texttt{pyMON} pipeline, show good overall agreement with those reported in the literature, and excellent agreement with studies that utilized similar amounts of data.


\subsection[Discussion of numax measurements]{Discussion of $\nu_{\rm max}$ measurements} \label{sec:numax-measurements}

\begin{table}[]
    \centering
    \caption{Summary of solar-type oscillation detections among the CDGs.}
    \begin{tabular}{lcc}
        \hline
        Detection Status & Number of CDGs & Percentage \\
        \hline
        \hline
        Clear detection & 43 & 33.3\% \\
        Marginal detection & 9 & 7.0\% \\
        No detection & 77 & 59.7\% \\
        \hline
    \end{tabular}
    \label{tab:oscillation_detections_statistics}
\end{table}


Table~\ref{tab:oscillation_detections_statistics} summarizes the detection statistics for solar-type oscillations among the CDGs. We found that 43 CDGs exhibit clear detections of solar-type oscillations, constituting $33.3\%$ of the sample, with 16 derived from $Kepler$ light curves and 27 from TESS light curves (see Table~\ref{tab:our-seismic-measurements}). In contrast, 9 CDGs ($7.0\%$) show marginal detections (see Table~\ref{tab:CDGs_marginal_detections}), while 77 CDGs ($59.7\%$) exhibit no detectable oscillations with the available data.  We note that this latter group includes many stars with limited data or sector coverage, so the true fraction of oscillating CDGs may be higher, and the $59.7\%$ should be considered an upper limit on the non-detection rate. The majority of the non-detections came from TESS light curves (see Table~\ref{tab:CDGs_no_detectable_oscillations} and Section~\ref{sec:numax-measurements}).


\begin{table*}
\centering
\begin{longtable}{lrcccccc}
\caption{Atmospheric parameters along with our asteroseismic measurements and mass estimates from equations (\ref{eq:mass3}) and (\ref{eq:mass5}) of the CDGs.}
\label{tab:our-seismic-measurements}\\
\hline

\multicolumn{1}{c}{Star} & KIC/TIC & $T_{\rm eff}$ & $\log g$  & $\log(L/L_{\odot})$ & $\nu_{\text{max}}$ & M$_{\text{seismic}}$ & M$_{\text{spec}}$ \\
     &     &  (K) &  & & ($\mu$Hz) & (M$_{\odot}$) & (M$_{\odot}$) \\
\hline
\hline
\endfirsthead

\multicolumn{8}{c}%
{{\bfseries \tablename\ \thetable{} -- continued from previous page}} \\
\hline

\multicolumn{1}{c}{Star} & KIC/TIC & $T_{\rm eff}$ & $\log g$  & $\log(L/L_{\odot})$ & $\nu_{\text{max}}$ & M$_{\text{seismic}}$ & M$_{\text{spec}}$ \\
     &     &  (K) &  & & ($\mu$Hz) & (M$_{\odot}$) & (M$_{\odot}$) \\
\hline
\hline
\endhead

\hline 
\endfoot

\hline
\endlastfoot

2M19581582+4055411 & 5737930 & 4443$\pm$50 & 1.78$\pm$0.05 & 2.77$\pm$0.07 & 3.61$\pm$0.17 & 1.72$\pm$0.30 & 3.67$\pm$0.75 \\
2M19055092+3745351 & 2423824 & 5007$\pm$50 & 2.27$\pm$0.05 & 2.15$\pm$0.05 & 21.79$\pm$0.12 & 1.64$\pm$0.20 & 1.69$\pm$0.28 \\
2M19252454+4036484 & 5446927 & 5107$\pm$50 & 2.30$\pm$0.05 & 2.05$\pm$0.05 & 21.97$\pm$0.10 & 1.22$\pm$0.15 & 1.33$\pm$0.22 \\
2M19090355+4407005 & 8222189 & 4914$\pm$50 & 2.37$\pm$0.05 & 2.17$\pm$0.04 & 22.68$\pm$0.36 & 1.91$\pm$0.19 & 2.40$\pm$0.37 \\
2M19211488+3959431 & 4830861 & 4959$\pm$50 & 2.46$\pm$0.05 & 2.14$\pm$0.05 & 26.02$\pm$0.14 & 1.98$\pm$0.20 & 2.65$\pm$0.41 \\
2M19382715+3827580 & 3355015 & 4846$\pm$50 & 2.35$\pm$0.05 & 1.81$\pm$0.05 & 26.79$\pm$0.13 & 1.03$\pm$0.13 & 1.06$\pm$0.18 \\
2M19442885+4354544 & 8110538 & 4975$\pm$50 & 2.25$\pm$0.05 & 1.98$\pm$0.05 & 28.76$\pm$0.15 & 1.49$\pm$0.18 & 1.12$\pm$0.19 \\
2M19400612+3907470 & 4071012 & 4992$\pm$50 & 2.56$\pm$0.05 & 2.07$\pm$0.04 & 29.21$\pm$0.14 & 1.85$\pm$0.18 & 2.77$\pm$0.42 \\
2M19340082+4108491 & 5881715 & 4840$\pm$50 & 2.37$\pm$0.05 & 1.89$\pm$0.05 & 29.97$\pm$0.24 & 1.39$\pm$0.17 & 1.34$\pm$0.22 \\
2M19422093+5018436 & 11971123 & 4848$\pm$50 & 2.48$\pm$0.05 & 1.79$\pm$0.04 & 31.87$\pm$0.24 & 1.17$\pm$0.12 & 1.29$\pm$0.20 \\
2M19404764+3942376 & 4667911 & 4740$\pm$50 & 2.42$\pm$0.05 & 1.67$\pm$0.04 & 35.33$\pm$0.24 & 1.07$\pm$0.11 & 0.98$\pm$0.15 \\
2M19133911+4011046 & 5000307 & 5018$\pm$50 & 2.62$\pm$0.05 & 1.75$\pm$0.04 & 41.85$\pm$0.18 & 1.24$\pm$0.12 & 1.46$\pm$0.22 \\
2M19181645+4506527 & 8879518 & 4832$\pm$50 & 2.72$\pm$0.05 & 1.73$\pm$0.04 & 45.21$\pm$0.22 & 1.46$\pm$0.15 & 2.08$\pm$0.32 \\
2M19125144+3850261 & 3736289 & 4978$\pm$50 & 2.58$\pm$0.05 & 1.80$\pm$0.04 & 49.09$\pm$0.24 & 1.68$\pm$0.17 & 1.59$\pm$0.24 \\
2M19565550+4330561 & 7848354 & 5004$\pm$50 & 2.66$\pm$0.05 & 1.75$\pm$0.06 & 50.74$\pm$0.29 & 1.52$\pm$0.22 & 1.65$\pm$0.30 \\
2M19052312+4422242 & 8352953 & 5101$\pm$50 & 3.34$\pm$0.05 & 1.10$\pm$0.04 & 230.13$\pm$0.40 & 1.45$\pm$0.14 & 1.64$\pm$0.25 \\
\hline
2M16405512+6435204 & 198279370 & 4591$\pm$50 & 2.02$\pm$0.05 & 2.83$\pm$0.04 & 5.67$\pm$0.18 & 2.76$\pm$0.29 & 6.44$\pm$0.72 \\
2M07263003+4230102 & 67885101 & 4766$\pm$50 & 2.14$\pm$0.05 & 2.30$\pm$0.05 & 11.11$\pm$0.34 & 1.40$\pm$0.18 & 2.15$\pm$0.28 \\
2M05272526+0317520 & 457250848 & 5057$\pm$50 & 2.04$\pm$0.05 & 2.45$\pm$0.07 & 11.23$\pm$0.34 & 1.63$\pm$0.27 & 1.90$\pm$0.34 \\
HD 91622$^{*}$ & 392847860 & 4457$\pm$50 & 1.63$\pm$0.10 & 2.27$\pm$0.05 & 11.42$\pm$0.20 & 1.70$\pm$0.21 & 0.81$\pm$0.21 \\
2M05485828-0335586 & 176582672 & 4614$\pm$50 & 1.95$\pm$0.05 & 2.10$\pm$0.06 & 12.62$\pm$0.51 & 1.13$\pm$0.17 & 1.00$\pm$0.15 \\
HD 102851 & 94550436 & 4991$\pm$41 & 2.68$\pm$0.26 & 2.42$\pm$0.04 & 13.31$\pm$0.34 & 1.88$\pm$0.19 & 8.18$\pm$4.96 \\
2M04024317+1638559 & 243013051 & 4826$\pm$50 & 2.26$\pm$0.05 & 2.35$\pm$0.05 & 13.66$\pm$0.09 & 1.85$\pm$0.22 & 3.03$\pm$0.40 \\
2M05301788+0140466 & 138827843 & 5130$\pm$50 & 2.44$\pm$0.05 & 1.93$\pm$0.06 & 14.70$\pm$0.34 & 0.61$\pm$0.09 & 1.36$\pm$0.22 \\
HD 91805 & 146431583 & 5247$\pm$15 & 2.56$\pm$0.04 & 2.46$\pm$0.04 & 15.19$\pm$0.47 & 1.98$\pm$0.19 & 5.59$\pm$0.73 \\
HD 49960 & 52980877 & 5030$\pm$32 & 2.61$\pm$0.05 & 2.28$\pm$0.06 & 16.32$\pm$0.51 & 1.63$\pm$0.23 & 4.89$\pm$0.89 \\
2M05120630-5913438 & 358459098 & 4998$\pm$50 & 2.35$\pm$0.05 & 1.80$\pm$0.04 & 18.95$\pm$0.42 & 0.64$\pm$0.07 & 0.92$\pm$0.10 \\
HD 31869 & 738036 & 4800$\pm$100 & 1.80$\pm$0.18 & 2.30$\pm$0.05 & 19.33$\pm$0.51 & 2.38$\pm$0.33 & 0.96$\pm$0.42 \\
2M00230981+7152126 & 363760185 & 4823$\pm$50 & 2.26$\pm$0.05 & 2.07$\pm$0.05 & 20.60$\pm$0.34 & 1.47$\pm$0.18 & 1.59$\pm$0.21 \\
BD+5 593 & 283623989 & 5045$\pm$60 & 2.50$\pm$0.18 & 2.08$\pm$0.05 & 20.95$\pm$0.51 & 1.31$\pm$0.16 & 2.37$\pm$1.03 \\
HD 78146 & 37923103 & 4734$\pm$96 & 2.13$\pm$0.03 & 2.27$\pm$0.05 & 22.11$\pm$0.51 & 2.67$\pm$0.37 & 2.02$\pm$0.32 \\
2M06022767-6209038 & 149989441 & 4946$\pm$50 & 2.40$\pm$0.05 & 2.11$\pm$0.04 & 23.00$\pm$0.77 & 1.65$\pm$0.17 & 2.19$\pm$0.55 \\
HD 56438 & 134545196 & 5037$\pm$163 & 2.75$\pm$0.01 & 2.19$\pm$0.05 & 23.03$\pm$0.86 & 1.86$\pm$0.31 & 5.47$\pm$0.96 \\
HD 18474 & 192247771 & 5198$\pm$38 & 2.65$\pm$0.03 & 2.15$\pm$0.04 & 25.29$\pm$0.48 & 1.67$\pm$0.16 & 3.50$\pm$0.42 \\
HD 82595 & 5480307 & 4995$\pm$79 & 2.28$\pm$0.01 & 2.35$\pm$0.04 & 25.58$\pm$0.51 & 3.07$\pm$0.34 & 2.76$\pm$0.32 \\
HD 94956 & 363418244 & 5131$\pm$75 & 2.76$\pm$0.30 & 2.16$\pm$0.05 & 26.27$\pm$0.86 & 1.86$\pm$0.24 & 4.85$\pm$3.41 \\
HD 201557 & 231638013 & 4730$\pm$90 & 2.15$\pm$0.18 & 2.19$\pm$0.05 & 31.74$\pm$0.30 & 3.19$\pm$0.43 & 1.77$\pm$0.77 \\
HD 18636 & 321087542 & 5085$\pm$80 & 2.70$\pm$0.18 & 1.68$\pm$0.04 & 32.52$\pm$0.69 & 0.78$\pm$0.09 & 1.45$\pm$0.62 \\
HD 54627 & 134282943 & 4990$\pm$70 & 2.55$\pm$0.16 & 2.06$\pm$0.05 & 33.33$\pm$1.20 & 2.06$\pm$0.27 & 2.65$\pm$1.03 \\
HD 124721 & 242443733 & 5107$\pm$61 & 2.64$\pm$0.11 & 2.13$\pm$0.04 & 34.72$\pm$1.03 & 2.33$\pm$0.25 & 3.50$\pm$0.96 \\
HD 16424 & 441127020 & 4850$\pm$40 & 2.60$\pm$0.10 & 1.95$\pm$0.04 & 37.27$\pm$0.34 & 1.98$\pm$0.19 & 2.58$\pm$0.65 \\
HD 40402 & 153122373 & 5005$\pm$110 & 2.80$\pm$0.18 & 2.16$\pm$0.05 & 44.56$\pm$0.34 & 3.43$\pm$0.48 & 5.86$\pm$2.57 \\
HD 166208 & 332626441 & 5177$\pm$52 & 2.81$\pm$0.04 & 2.04$\pm$0.04 & 48.03$\pm$1.26 & 2.49$\pm$0.59 & 3.99$\pm$1.00 \\

\end{longtable}

\begin{minipage}{\textwidth}
\vspace{0.5em}
\small \textbf{Note:} All stellar designations beginning with `2M' denote the APOGEE ID of the carbon-deficient giants identified using APOGEE data. For these stars, we adopt the errors in $T_{\rm eff}$ and [Fe/H] as $50$~K and $0.05$~dex, respectively. The first 16 rows contain $Kepler$ Input Catalog (KIC) IDs, while the remaining rows contain TESS Input Catalog (TIC) IDs. \\ A horizontal line is used to visually separate stars observed with $Kepler$ from those observed with TESS.
$^{*}$ $T_{\rm eff}$ and $\log g$  are from \cite{Matsuno2024}.
\end{minipage}
\end{table*}

\begin{figure}[]
    \centering
    \includegraphics[width=\linewidth]{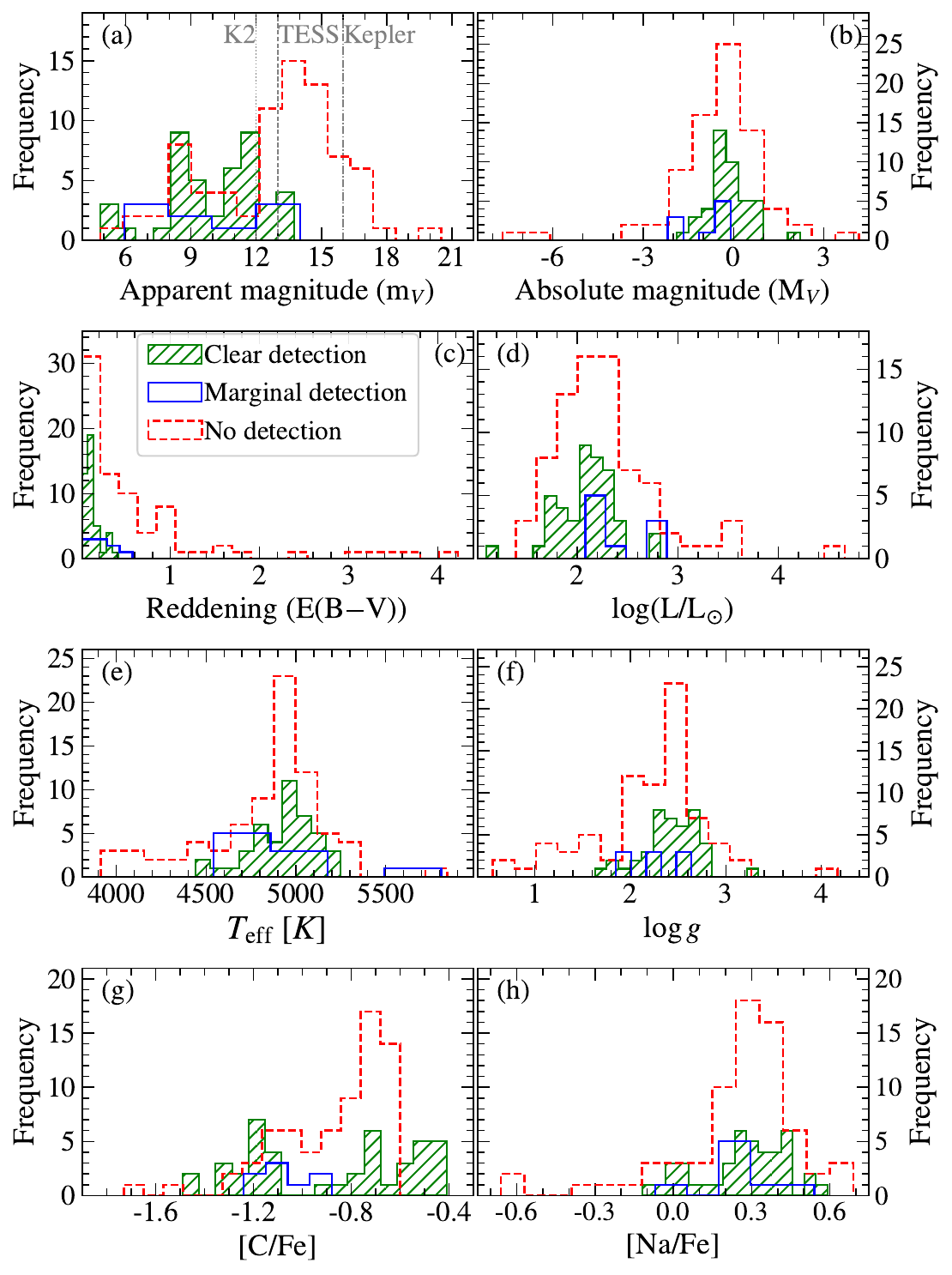}
    \caption{Histograms of various stellar parameters for CDGs, color-coded by $\nu_{\text{max}}$ detection level: green (clear), blue (marginal), and red (no detection). Panels (a) and (b) show the distributions of apparent and absolute magnitudes, respectively. In Panel (a), vertical lines indicate the  apparent magnitude detection limits for the K2, TESS, and $Kepler$ missions, respectively. Panels (c) and (d) display the distributions of reddening and luminosity. Panels (e) and (f) present the distributions of effective temperature and spectroscopic surface gravity. Finally, panels (g) and (h) illustrate the distributions of [C/Fe] and [Na/Fe] abundance ratios. We use the solar abundance of C and Na as derived by \cite{Grevesse2007}. The CDG HD~91622, which exhibits clear detections of solar-type oscillations, lacks literature values for $T_{\rm eff}$ and $\log g$ ; therefore, these parameters were taken from photometric measurements in \citet{Matsuno2024}. Notably, HD~91622 is not included in the distributions of [C/Fe] and [Na/Fe]. See text for details.}
    \label{fig:detection_levels_information}
\end{figure}

A key science question in this study is to deduce the origins of CDGs, which requires careful consideration of observational biases that may affect our sample. To investigate this, Figure~\ref{fig:detection_levels_information} presents histograms of various stellar parameters for CDGs, categorized by their $\nu_{\text{max}}$ detection levels: clear detection, marginal detection, and no detection. These distributions allow us to examine how different observational and intrinsic properties influence oscillation detectability and to determine whether biases exist that could affect our conclusions about CDG origins.

Figure~\ref{fig:detection_levels_information}(a) shows apparent magnitudes, estimated from color-color transformations in \cite{Riello2021} relating $Gaia$~DR3 to Johnson-Cousins photometry \citep{Stetson2000}. CDGs with clear $\nu_{\text{max}}$ detections are brighter/nearby stars, typically m$_{V} \leq 12$. Panel (b) shows absolute magnitudes. CDGs with clear detections cluster around M$_{V} \approx -0.2$, whereas stars with no detections exhibit a broader range of absolute magnitudes. This suggests the non-detected sample is not biased toward specific magnitudes. The similarity of the distribution to the overall sample implies no strong evolutionary bias in $\nu_{\text{max}}$ detection. Panel (d) shows luminosities, determined as in Section~\ref{sec:luminosities}. CDGs with clear detections have $\log(L/L_{\odot}) \approx 2.1 \pm 0.3$~dex. Non-detections occur at high luminosity because these stars exhibit lower oscillation frequencies (low $\nu_{\text{max}}$), which are hard to resolve with a light curve of limited duration (insufficient frequency resolution).  Indeed, the majority of non-detections are based on TESS data (there was only one from $Kepler$ and two from K2). This reflects the fact that the observational limitation is the SNR, which significantly influences  detectability \citep[e.g.,][]{Stello2022}. In particular, fainter and intrinsically less luminous stars often have an SNR inadequate for detecting their oscillations \citep{Stello2022}. Panel (c) shows lower reddening for clear detections, from the 3D dust map of \cite{Green2019} or, where unavailable, from \cite{Schlegel1998} recalibrated by \cite{Schlafly2011}. This is likely due to proximity.

Figure~\ref{fig:detection_levels_information}(e) and (f) show the distributions of effective temperature and surface gravity, respectively, with values taken from the literature \citep{Adamczak2013, Palacios2016, Holanda2023, Maben2023a, Maben2023b, Holanda2024}. Clear detections occur at $T_{\text{eff}} \approx 4900 \pm 200$~K and $\log g \approx 2.4 \pm 0.3$~dex. This is consistent with red clump stars expected at $\log g$  $\approx 2.4$ \citep{Yu2018}. No clear detections appear at cooler $T_{\text{eff}} \approx 4000-4500$~K or lower $\log g \approx 0.5-1.5$. This confirms that cooler, larger giants with low $\nu_{\text{max}}$ are underrepresented in the asteroseismic detections. This again reflects that fact that the detection of solar-like oscillations becomes increasingly difficult stars with low $\nu_{\text{max}}$, due to insufficient frequency resolution. Panels (g) and (h) of Figure~\ref{fig:detection_levels_information} present the distributions of [C/Fe] and [Na/Fe], respectively. Clear detections favor extreme [C/Fe] compositions. They also show a secondary bias toward moderate [C/Fe] compositions. The sodium abundance distribution remains similar between CDGs with and without clear detections.

In summary, our analysis of $\nu_{\text{max}}$ detection rates in CDGs reveals that successful detections are primarily limited by observational constraints, with clear biases toward brighter, warmer stars with properties consistent with red clump stars. The under-representation of cooler, more luminous giants with low $\nu_{\text{max}}$ values in our sample highlights the technical challenges in resolving low-frequency oscillations, particularly with TESS data, while the preference for carbon-poor compositions suggests a potential connection between stellar chemistry and oscillation properties.

\subsection[Comparison of expected and observed numax]{Comparison of expected and observed $\nu_{\rm max}$}  \label{sec:expected_vs_observed_numax}

\begin{figure}
    \centering
    \includegraphics[width=1\linewidth]{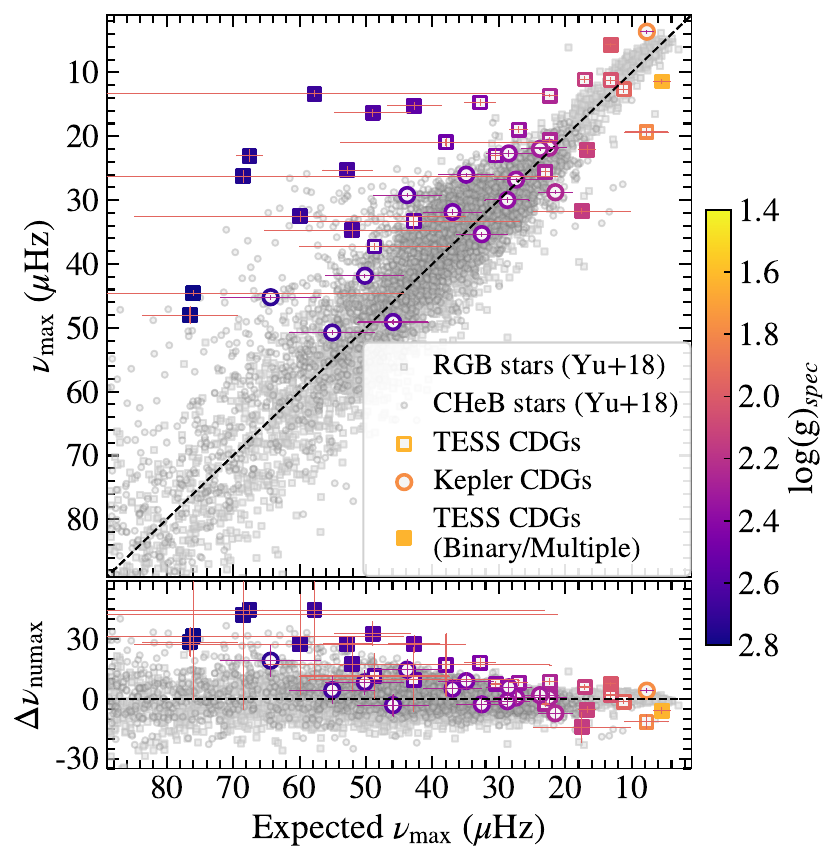}
    \caption{Comparison of the expected $\nu_{\text{max}}$ values (using Equation~(\ref{eq:initial_numax})) against the observed $\nu_{\text{max}}$ for the CDGs exhibiting clear detections of solar-type oscillations. Giants classified based on asteroseismic analysis form the background from \cite{Yu2018} (see key). Their expected $\nu_{\text{max}}$ values were calculated using $T_{\rm eff}$ and $\log g$ values taken from APOGEE~DR17 \citep{Abdurrouf2022}. The color bar indicates the spectroscopic surface gravity of each CDG. Large open circles represent our $Kepler$ CDGs, while open squares indicate our TESS CDGs. The CDGs that are reported to be in multiple systems or in a binary system in the literature are filled with colors corresponding to their spectroscopic surface gravity. KIC~8352953, which is classified as an RGB star in \cite{Yu2018} and has a $\nu_{\text{max}}$ $\approx 230$~$\mu$Hz, corresponding to the lower-RGB phase, is not included. The bottom panel shows the star-by-star residuals between expected $\nu_{\text{max}}$  and observed $\nu_{\text{max}}$. See text for details.}
    \label{fig:expected_vs_measured_numax}
\end{figure}

In Figure~\ref{fig:expected_vs_measured_numax} we compare the expected $\nu_{\text{max}}$, derived from spectroscopic values for $\log g$  and $T_{\rm eff}$ using Equation~(\ref{eq:initial_numax}), with the observed $\nu_{\text{max}}$ for our sample of CDGs, together with a large sample of RGB and RC stars from \citet{Yu2018}.

To identify binaries and multiple systems in our sample, we cross-matched the CDGs with several established catalogues: the Washington Visual Double Star catalogue \citep{Mason2001}, the Catalog of Components of Double and Multiple Stars \citep{Dommanget2002}, an extensive catalogue of $Gaia$~eDR3 binary stars within \(\approx 1\) kpc of the Sun\footnote{We note that $Gaia$ catalogue is of spatially resolved binaries only.} \citep{Badry2021}, the catalogue of stellar and substellar companions identified using $Gaia$~DR2 and EDR3 proper motion anomalies \citep{Kervella2019, Kervella2022}, and the Binary Star Database \citep{Kovaleva2015}. This analysis identified 14 out of 43 CDGs as binaries or multiple systems, which are marked with filled symbols in Figure~\ref{fig:expected_vs_measured_numax}.

Most single CDGs lie close to the one-to-one line, indicating good agreement between expected and observed $\nu_{\text{max}}$. In contrast, binaries and multiple systems display the largest discrepancies, with observed $\nu_{\text{max}}$ values systematically lower than expected. This trend is especially clear in the residuals shown in the bottom panel of Figure~\ref{fig:expected_vs_measured_numax}. The larger horizontal error bars for binaries reflect greater uncertainties in spectroscopic $\log g$ and $T_{\rm eff}$, which further complicate the prediction of $\nu_{\text{max}}$ for these systems.

For single stars, we find that the median residual, defined as the difference between the expected and observed $\nu_{\text{max}}$, is $5.1~\mu$Hz when using APOGEE spectroscopic parameters and $27.4~\mu$Hz when using optical spectroscopy, indicating larger systematics in the latter (see Section~\ref{sec:seismic_logg}). Small changes in $\log g$ strongly affect $\nu_{\text{max}}$ through the scaling relation. A typical uncertainty for spectroscopic $\log g$ is $0.1$~dex. Applying a $-0.1$~dex offset reduces $g$ by 20.6\%, causing an equivalent decrease in $\nu_{\text{max}}$ for fixed $T_{\rm eff}$. For example, for HD~56438, with $\log g = 2.75$ and $T_{\rm eff} = 5037$~K \citep{Palacios2016}, this lowers the expected $\nu_{\text{max}}$ from $67.6~\mu$Hz to $53.7~\mu$Hz, reducing the residual from the observed $23.0~\mu$Hz. Systematic errors, such as those from peculiar abundances \citep{Palacios2016}, can further bias $\log g$ and $T_{\rm eff}$, increasing deviations.

These results suggest that contamination from unresolved companions in binaries likely biases spectroscopic measurements, leading to erroneous predictions for $\nu_{\text{max}}$. In summary, agreement between expected and observed $\nu_{\text{max}}$ is generally good for single CDGs, but breaks down for binaries and multiple systems, where spectroscopic measurements of $\log g$ and/or $T_{\rm eff}$ predict values of $\nu_{\text{max}}$ that are greater than we observe.

\subsection{Determination of stellar parameters}
\subsubsection{Luminosities} \label{sec:luminosities}
Luminosities of the CDGs were determined using the standard formula:
\begin{equation}
\log(L/L_{\odot}) = -0.4 [V_{0} - 5\log(d/\text{pc}) + BC - M_{\rm bol,\odot}].
\end{equation}

Distances ($d$) were taken from the catalog of \cite{Bailer-Jones2021}. The visual magnitudes ($V_0$) and their errors were estimated from the color–color transformations provided in \cite{Riello2021}, which relate the $Gaia$~DR3 photometric system to the Johnson-Cousins system \citep{Stetson2000}. We applied bolometric corrections (BC) to the absolute magnitudes following the relation from \cite{Alonso1999}. Reddening values were taken from the three-dimensional dust map of \cite{Green2019}. In cases where the reddening was not available, we took them from the \cite{Schlegel1998} map which is re-calibrated in \cite{Schlafly2011}. Our derived luminosities are listed in Table~\ref{tab:our-seismic-measurements}.

\begin{figure*}[]
    \centering
    \includegraphics[width=\linewidth]{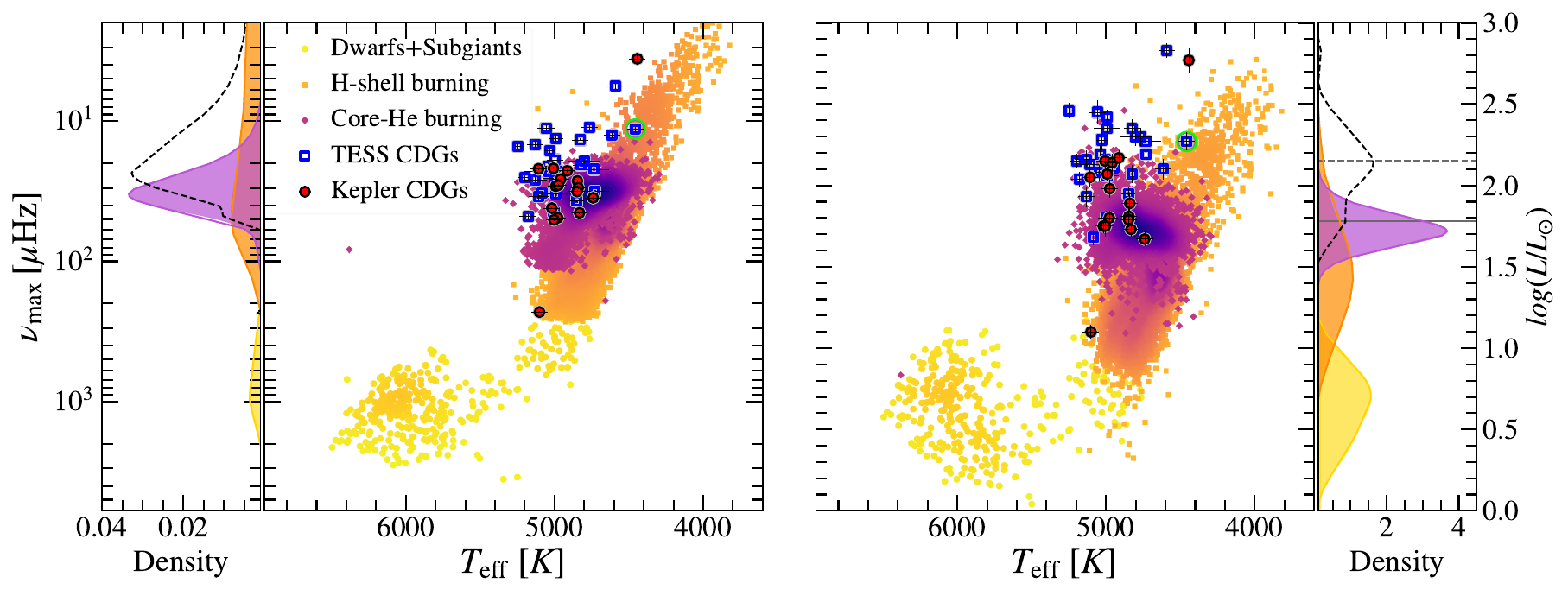}
    \caption{Left panel: Asteroseismic Hertzsprung–Russell diagram ($\nu_{\text{max}}$ versus $T_{\rm eff}$). Right panel: Hertzsprung–Russell diagram. The $\nu_{\text{max}}$ estimates were adopted from \cite{Serenelli2017} for the dwarfs and subgiants (yellow symbols), \cite{Pinsonneault2018} and \cite{Yu2018} for the first-ascent RGB stars (orange symbols), and He-core burning giants (purple symbols). Their estimates of $T_{\rm eff}$ were taken from APOGEE~DR17 \citep{Abdurrouf2022}, and luminosity estimates were taken from the $Gaia$~DR2 catalog \citep{Gaia2018}. The large filled red circles and open blue squares show our $Kepler$ and TESS CDGs, respectively. The CDG highlighted with a green circle is HD~91622, which lacks $T_{\rm eff}$ (and carbon abundance) from high-resolution spectra; its $T_{\rm eff}$ was adopted from \cite{Matsuno2024}. Kernel density histograms are included to show the distributions for various parts of the sample, with the color scheme following the Hertzsprung–Russell diagram, and the dashed histogram showing our CDG sample. Horizontal lines mark CDG luminosity peaks: a solid line at $\log(L/L_{\odot}) = 1.78$~dex aligns with the core He-burning phase for low-mass stars (i.e., $\log (L/L_{\odot}) \simeq 1.55- 1.85$~dex; \citealt{Girardi2016}), and a dashed line at $\log(L/L_{\odot}) = 2.15$~dex marks the primary peak where most CDGs lie.}
    \label{fig:HRD_main}
\end{figure*}

The right panel of Figure~\ref{fig:HRD_main} shows the Hertzsprung-Russell diagram (log($L/L_{\odot}$) versus $T_{\rm eff}$). With the exception of KIC~8352953, the CDGs exhibit luminosities characteristic of RC or early asymptotic giant branch (EAGB) stars. Their luminosity distribution features a primary peak at $\log(L/L_{\odot}) = 2.15$~dex and a secondary peak at $\log(L/L_{\odot}) = 1.78$~dex, which correspond to the two luminosity groups in the bimodal distribution reported by \cite{Maben2023b}, followed by a bright tail extending towards the RGB tip (see also Table~\ref{tab:our-seismic-measurements}). The secondary peak aligns with core He-burning predictions for low-mass stars (i.e., $\log (L/L_{\odot}) \simeq 1.55- 1.85$~dex; \citealt{Girardi2016}). KIC~8352953, which is also included in the \cite{Maben2023b} study and classified as an RGB star by \citet{Yu2018}, aligns with the lower RGB.

We also include an asteroseismic Hertzsprung–Russell diagram in the left panel of Figure~\ref{fig:HRD_main}. Most CDGs show $\nu_{\text{max}}$ values lower than typical RC stars, extending to even lower frequencies in the tail, consistent with post-RC evolution, as reflected in the luminosity distribution in the right panel of Figure~\ref{fig:HRD_main}. Notably, the more luminous CDGs ($\log(L/L_{\odot}) \gtrsim 2.15$~dex) possess higher $T_{\rm eff}$ than typical RGB stars at similar luminosities, reinforcing their interpretation as bright RC or EAGB stars \citep{Maben2023b} rather than RGB stars. This bright tail comprises a mixture of higher-mass CDGs (see Section~\ref{sec:group2a}) and stars that could be either undergoing post-RC evolution towards the EAGB phase, which runs parallel to the RGB and lasts longer than the later stages of the AGB phase, or be massive post-merger core He-burning stars with larger-than-normal helium cores. Only KIC~8352953 exhibits a $T_{\rm eff}$ consistent with the region where RGB stars are typically found, as evident in both panels.

\begin{figure}
    \centering
    \includegraphics[width=1.0\linewidth]{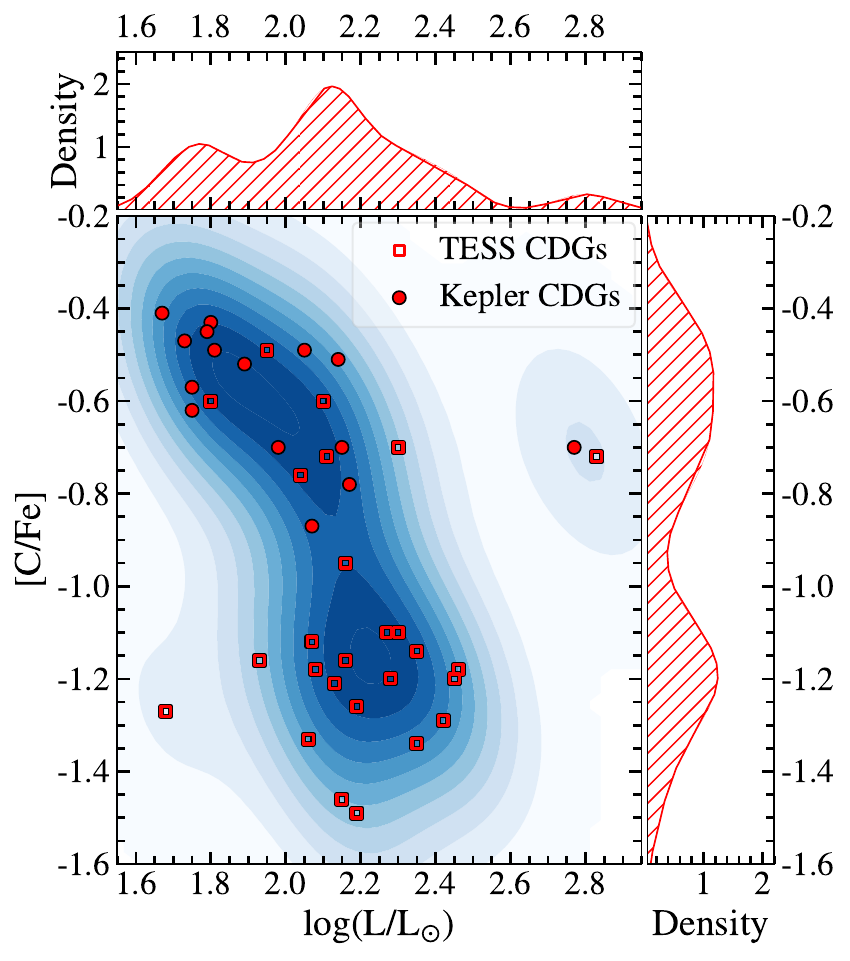}
    \caption{Carbon abundance as a function of $\log \rm(L/L_{\odot})$ for the CDGs for which $\nu_{\text{max}}$ has been estimated, excluding KIC~8352953, which is a lower-luminosity RGB star, and HD~91622, which lacks carbon abundance data from high-resolution spectra. Contours represent the density distribution. Kernel density histograms illustrate the distributions of the CDGs. Solar carbon abundance is from \cite{Grevesse2007}.}
    \label{fig:luminosity_vs_carbon}
\end{figure}

Figure~\ref{fig:luminosity_vs_carbon} shows carbon abundance versus luminosity for the CDG sample, revealing a clear bimodality in the luminosity distribution. The two groups are centered at $\log(L/L_{\odot}) = 2.0 \pm 0.1$~dex and $2.2 \pm 0.1$~dex, broadly corresponding to the RC-like and EAGB-like populations in Figure~\ref{fig:HRD_main}. The more luminous group ($\log(L/L_{\odot}) \approx 2.2$~dex) exhibits significantly lower carbon abundances than the fainter group, indicating a carbon-luminosity anti-correlation. This trend is consistent with the results of \citet{Maben2023b}, who identified a similar bimodality in a smaller sample of 14 CDGs, with peaks at $\log(L/L_{\odot}) = 1.8 \pm 0.1$~dex and $2.1 \pm 0.1$~dex (see their Figure~3). Our larger sample of 41 stars provides stronger statistical support for this luminosity bimodality. Two CDGs with $\log(L/L_{\odot}) \approx 2.8$~dex, exceeding typical RC or EAGB luminosities, may represent a later, more luminous EAGB phase, or could be merger products with larger He-burning cores.

\subsubsection{Masses} \label{sec:masses}

\begin{figure}
    \centering
    \includegraphics[width=1.0\linewidth]{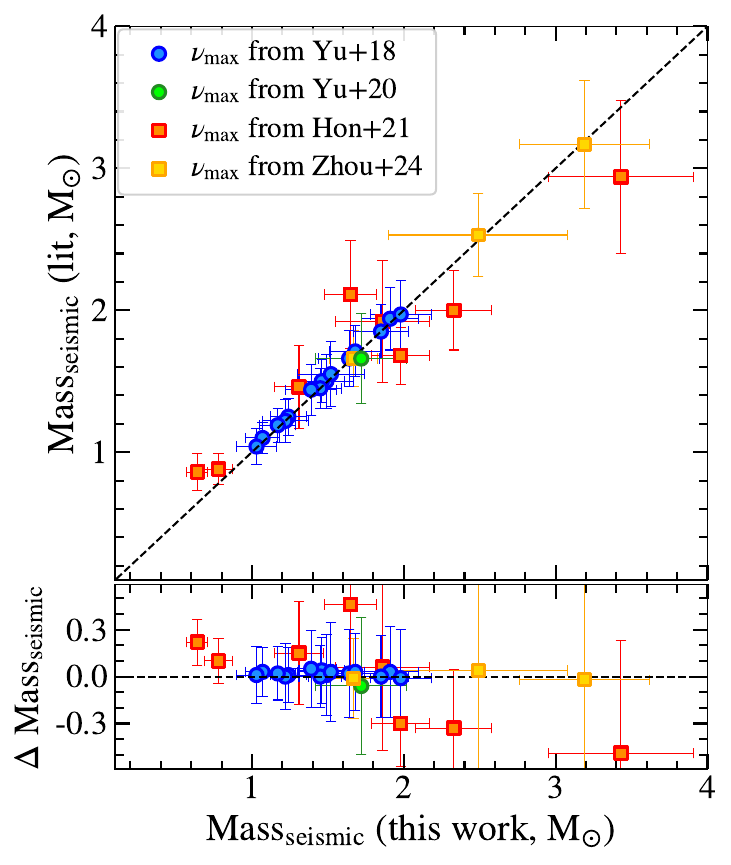}
    \caption{Comparison of seismic masses for 27 CDGs with published $\nu_{\mathrm{max}}$ estimates. Masses are derived using $\nu_{\mathrm{max}}$ values from this study (via the \texttt{pyMON} pipeline) versus published $\nu_{\mathrm{max}}$ values, both calculated with Equation~(\ref{eq:mass3}). The bottom panel shows star-by-star residuals. The agreement is excellent, apart from \citet{Hon2021} (limited data coverage; see Section~\ref{sec:comparision-with-published-numax}).}
\label{fig:literature_seismic_mass_comparision}
\end{figure}

Figure~\ref{fig:literature_seismic_mass_comparision} compares seismic masses for the 27 CDGs  derived from published $\nu_{\mathrm{max}}$ measurements \citep{Yu2018, Yu2020, Hon2021, Zhou2024} with those calculated in this study using $\nu_{\mathrm{max}}$ from the \texttt{pyMON} pipeline, both determined with Equation~(\ref{eq:mass3}).  Overall, the masses show excellent agreement. Although masses based on $\nu_{\mathrm{max}}$ values from \citet{Hon2021} are systematically lower by approximately 2.3\%, they agree with ours within their respective uncertainties. In contrast, masses derived from $\nu_{\mathrm{max}}$ values reported by \citet{Yu2018, Yu2020} and \citet{Zhou2024} agree with ours within 1\% (see Section~\ref{sec:comparision-with-published-numax} and Figure~\ref{fig:pyMON-numax-comparision}).

\begin{figure}
    \centering
        \centering
        \includegraphics[width=1.0\linewidth]{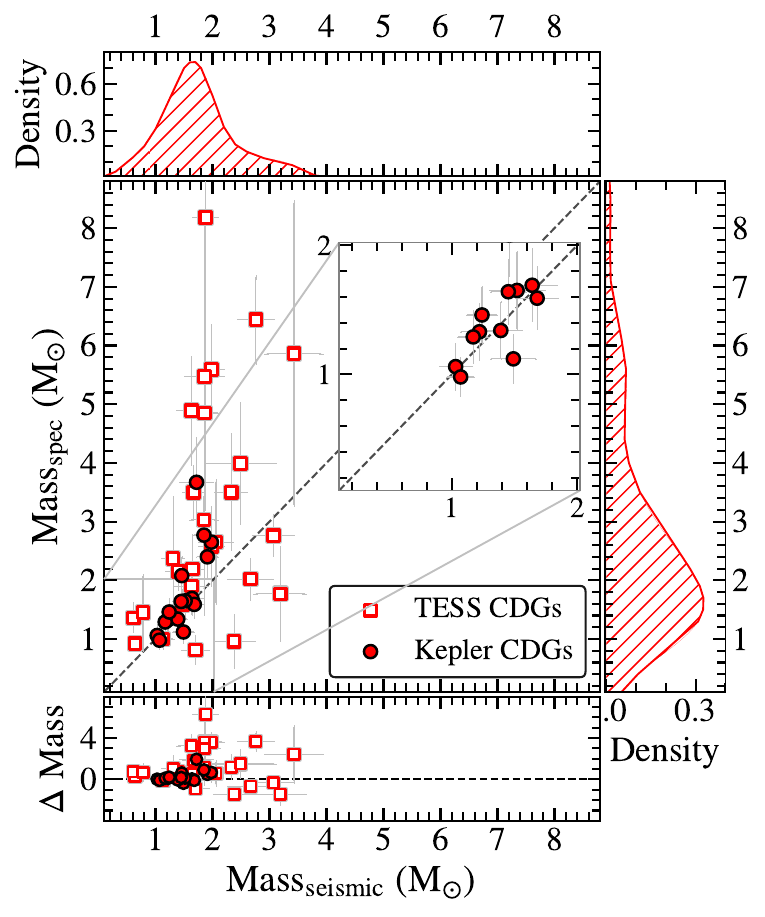}
    \caption{Comparison of seismic masses for the entire seismic CDG sample (43 stars) calculated using Equation~(\ref{eq:mass3}) with $\nu_{\text{max}}$ from the \texttt{pyMON} pipeline versus non-seismic masses from Equation~(\ref{eq:mass5}). The black dashed line again indicates the one-to-one relation, and kernel density histograms show the mass distributions. The zoomed inset highlights $Kepler$ stars. The bottom panel shows star-by-star residuals between non-seismic and seismic masses.}
    \label{fig:seismic_spec_masses_comparision}
\end{figure}

We present the seismic and non-seismic mass determinations for all 43 CDGs in Table~\ref{tab:our-seismic-measurements} and Figure~\ref{fig:seismic_spec_masses_comparision}. The non-seismic masses were calculated using
\begin{equation}
  \left( \frac{M}{M_\odot}\right)  \simeq \left(\frac{L}{L_{\odot}}\right) \left(\frac{g}{g_{\odot}}\right) \left(\frac{T_{\rm eff}}{T_{\rm eff,\odot}}\right)^{-4}.\label{eq:mass5}
\end{equation}

Figure~\ref{fig:seismic_spec_masses_comparision} extends the analysis to our entire CDG sample, comparing individual seismic and non-seismic masses. For $Kepler$ stars there is good agreement between the two methods for stars with  $M\leq 2~\mathrm{M}_{\odot}$, as shown in the zoomed inset. For higher-mass stars and those observed by TESS, the seismic and non-seismic methods show disagreement, with non-seismic masses generally exceeding their seismic counterparts.

As discussed in Section~\ref{sec:luminosities}, many of our CDGs are luminous giants with $\log(L/L_\odot) \gtrsim 2.15$. In this regime, classical asteroseismic scaling relations involving both $\nu_{\text{max}}$ and $\Delta\nu$ are known to overestimate masses unless corrected (e.g., with $f_{\Delta\nu}$ factors). However, our mass estimates rely only on $\nu_{\text{max}}$, and recent work has shown that the $\nu_{\text{max}}$-only scaling relation performs more reliably at high luminosity without requiring such corrections \citep{Howell2024,Ash2025}.

The mode of the seismic mass distribution for the CDGs is $1.70~M_\odot$ (see top panel of Figure~\ref{fig:seismic_spec_masses_comparision}), with a spread ranging from $0.61~M_\odot$ to $3.43~M_\odot$. In contrast, the mode of the non-seismic mass distribution is $1.59~M_\odot$ (see right panel of Figure~\ref{fig:seismic_spec_masses_comparision}), with a broader spread from $0.81~M_\odot$ to $8.18~M_\odot$. While both mass distributions peak at similar values, the seismic masses exhibit a much narrower spread. This difference is also reflected in the typical uncertainties: the mean uncertainty for the seismic masses is $0.21~M_\odot$, whereas for the non-seismic masses it is substantially larger at $0.67~M_\odot$.

\subsubsection{Seismic surface gravities} \label{sec:seismic_logg}
We estimated the seismic surface gravities of the CDGs using the scaling relation:
\begin{equation}
\left( \frac{g}{g_{\odot}} \right) \simeq \left( \frac{\nu_{\text{max}}}{\nu_{\text{max},\odot}} \right) \left( \frac{T_{\rm eff}}{T_{\rm eff,\odot}} \right)^{-1/2}.
\label{eq:seismic_surface_gravity}
\end{equation}

Asteroseismic $\log g$ uncertainties average $\pm0.01$~dex, far lower than spectroscopic $\log g$ uncertainties of $\pm0.07$~dex from high-resolution spectra. The offset between spectroscopic and seismic $\log g$ is \[\log g_{\text{spec}} - \log g_{\text{seis}} = 0.06 \pm 0.22 \, \text{dex,}\] which is consistent with zero within the uncertainties. However, as discussed in Section~\ref{sec:expected_vs_observed_numax}, standard solar-scaled models can yield incorrect spectroscopic $\log g$ values for CDGs due to their peculiar chemistry \citep{Palacios2016}.

\begin{figure}[]
    \centering
    \includegraphics[width=0.9\linewidth]{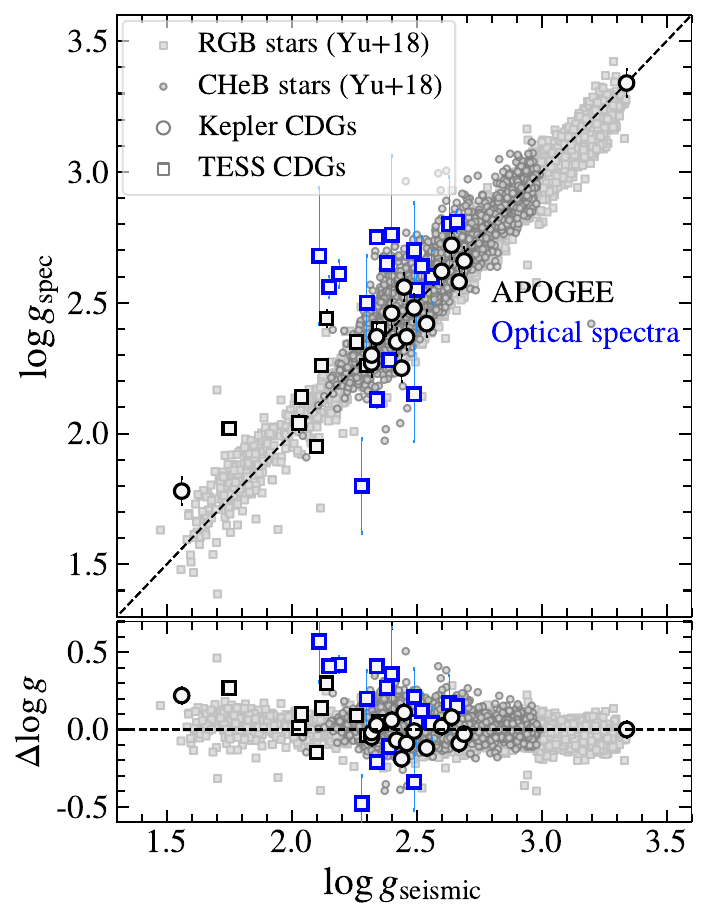}
     \caption{Comparison of seismic surface gravities using $\nu_{\text{max}}$ values from this study versus spectroscopic surface gravities for CDGs from literature, excluding HD~91622, since it lacks spectroscopic $\log g$ measurement. Background RGB and RC stars from \citet{Yu2018} use spectroscopic $\log g$ from APOGEE~DR17, with seismic $\log g$ derived using \citet{Yu2018} $\nu_{\text{max}}$ and APOGEE~DR17 $T_{\rm eff}$. The black dashed line denotes the one-to-one relation. CDGs with spectroscopic $\log g$ from APOGEE~DR17 (black) and optical spectra (blue) are also shown. The bottom panel shows star-by-star residuals between spectroscopic and seismic $\log g$.}
    \label{fig:surface_gravity_comparision}
\end{figure}

Figure~\ref{fig:surface_gravity_comparision} compares these values, showing spectroscopic $\log g$ from APOGEE~DR17 and optical spectra against seismic $\log g$, with RGB and RC stars from \citet{Yu2018} as background. The systematic offsets are quantitatively smaller for APOGEE-derived measurements ($0.02 \pm 0.12$~dex) than optical spectra ($0.13 \pm 0.28$~dex), due to the application of a neural network-based calibration in APOGEE that incorporates asteroseismic parameters \citep{Holtzman2018, Abdurrouf2022}. For \citet{Yu2018} giants, spectroscopic $\log g$ exceeds seismic values at lower $\log g$, especially for RC stars, consistent with APOGEE trends \citep{Pinsonneault2014}. These differences further support our analysis in Section~\ref{sec:expected_vs_observed_numax} regarding the challenges in spectroscopic $\log g$ determinations for giants \citep{Masseron2017}. The superior accuracy of seismic $\log g$ determinations significantly enhances the reliability of evolutionary classifications.

\subsection{Characterization of carbon-deficient stars}
\subsubsection{Three groups of CDGs}

In the previous study of 15 $Kepler$ field CDGs \citep{Maben2023b}, 14 were RC stars. Based on available data, we divided them into two groups based on their luminosity and chemical patterns: Group~1 CDGs (normal-luminosity RC stars; 8 stars); and Group~2 CDGs (over-luminous RC stars; 6 stars). The remaining CDG was an RGB star. We also compared these CDGs to 29 previously known wGb stars, and also assigned them to groups, based on their chemistry and luminosities: Group~3a (6 wGb stars) and Group 3b (23 wGb stars). Group 3a consisted of  primarily low-mass stars, with limited seismic data from TESS, but which appeared overluminous for their masses and showed evidence of more extreme chemical pollution as compared to Groups 1 and 2. Group~3b were wGb stars without any seismic data, which also showed extreme chemical pollution but lacked seismic mass constraints.

\begin{figure*}[]
    \centering
    \includegraphics[width=0.9\linewidth]{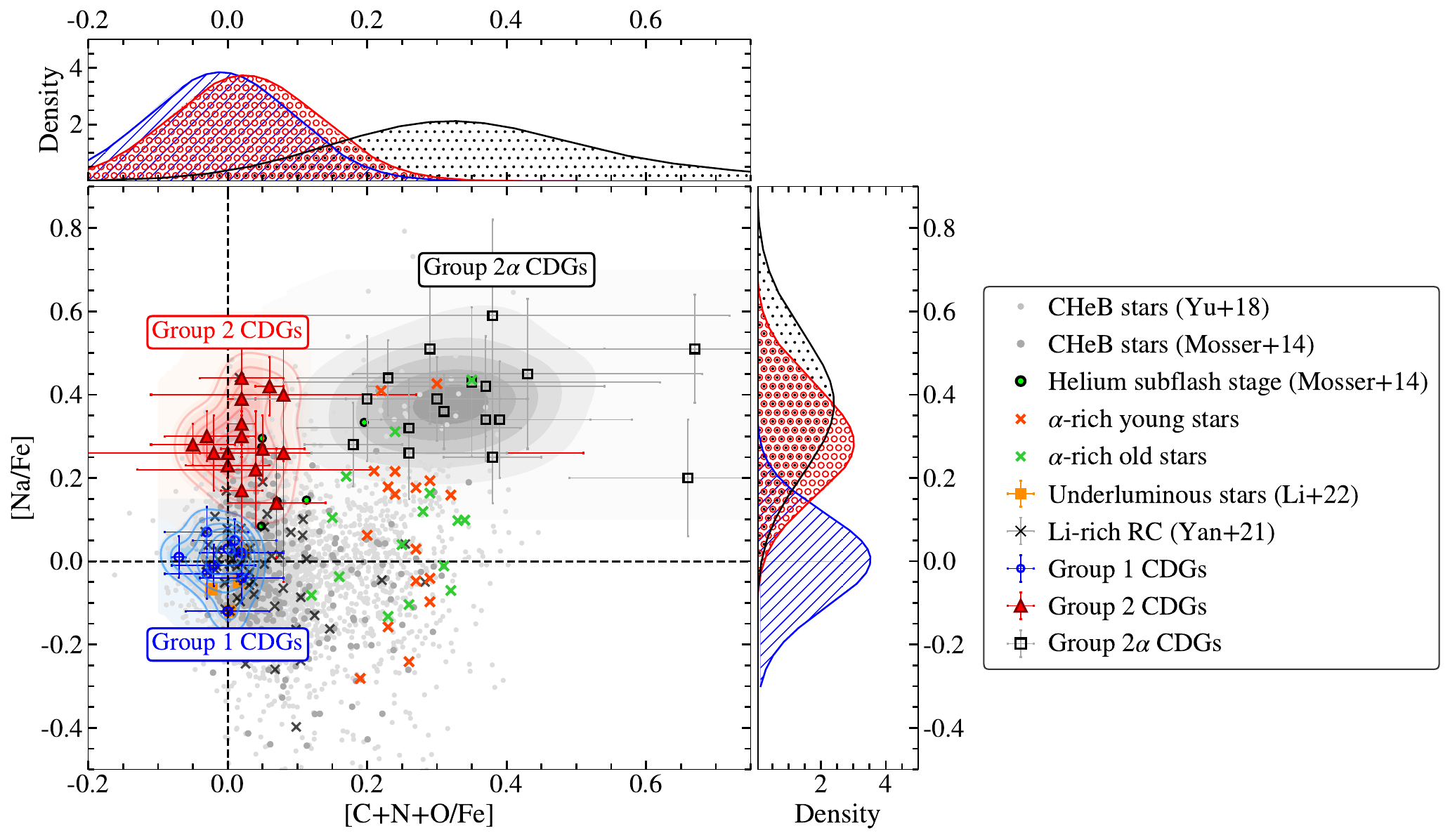}
    \caption{Trends of sodium abundance vs. the [C+N+O/Fe] abundance ratio. Giants classified based on asteroseismic analysis form the background from \cite{Mosser2014} and \cite{Yu2018} (small filled circles; see key). The underluminous stars from \cite{Li2022} are shown by orange squares. The abundance ratios of the normal giants and the underluminous stars are from APOGEE~DR17. The Group~1, Group~2, and Group~2$\alpha$ CDGs are represented by blue circles, red triangles, and black squares, respectively. The Li-rich giants (black crosses) are from the \cite{Yan2021} study. These Li-rich giants are the low-spectral-resolution RC sample that had good-quality APOGEE~DR17 data (59 stars). The $\alpha$-rich young (orange crosses) and old stars (green crosses) were compiled by \citet{Hekker2019} from the \citet{Jofre2016} and \citet{Pinsonneault2018} studies. Solar abundances are from \cite{Grevesse2007}. Contours highlight the separation into three groups based on their Na and CNO abundances.}
    \label{fig:sodium_vs_CNO}
    \end{figure*}

In this study our sample with seismic data (43 CDGs) is approximately three times the size of the previous sample. In this sample we find that while there is a clear progression from lower to higher luminosities between the groups, significant overlap in luminosity among the CDGs makes it challenging to define distinct groups based solely on this parameter (see Figures~\ref{fig:HRD_main} and~\ref{fig:luminosity_vs_carbon}). A clearer distinction emerges in the [Na/Fe] versus [C+N+O/Fe] abundance plane (see Figure~\ref{fig:sodium_vs_CNO}). Here, the CDGs separate into three distinct groups based on their chemical abundances:
\begin{itemize}[itemsep=1pt,topsep=1pt]
    \item Group~1 CDGs: Normal [Na/Fe] abundances with scaled-solar [C+N+O/Fe] compositions.
    \item Group~2 CDGs: Enhanced [Na/Fe] abundances, also with scaled-solar [C+N+O/Fe] compositions.
    \item Group~$2\alpha$ CDGs: Enhanced in both [Na/Fe] and [C+N+O/Fe], with an average [C+N+O/Fe] enhancement of $+0.4$~dex. This group aligns with the previously identified wGb stars, reinforcing its distinct nature.
\end{itemize}

The naming of the third group as 2$\alpha$ is based on further inspection which indicates that it is very similar to Group 2, as we describe below.

\subsubsection{Deciphering Group 2$\alpha$}
\label{sec:group2a}
In Table~\ref{tab:CDG_groups_comparison}
we collate and summarise all available information from the current study, including atmospheric parameters, abundances, masses, and luminosities. Here we see that Groups 1 and 2 are quite distinct,
with effective temperatures, luminosities, and masses increasing between them. Chemical abundances also show progressively higher [Na/Fe], greater [C/Fe] depletion, and a decreasing [C/N] ratio. The differences between these groups suggest distinct formation pathways.

\begin{figure}
    \centering
    \includegraphics[width=1.0\linewidth]{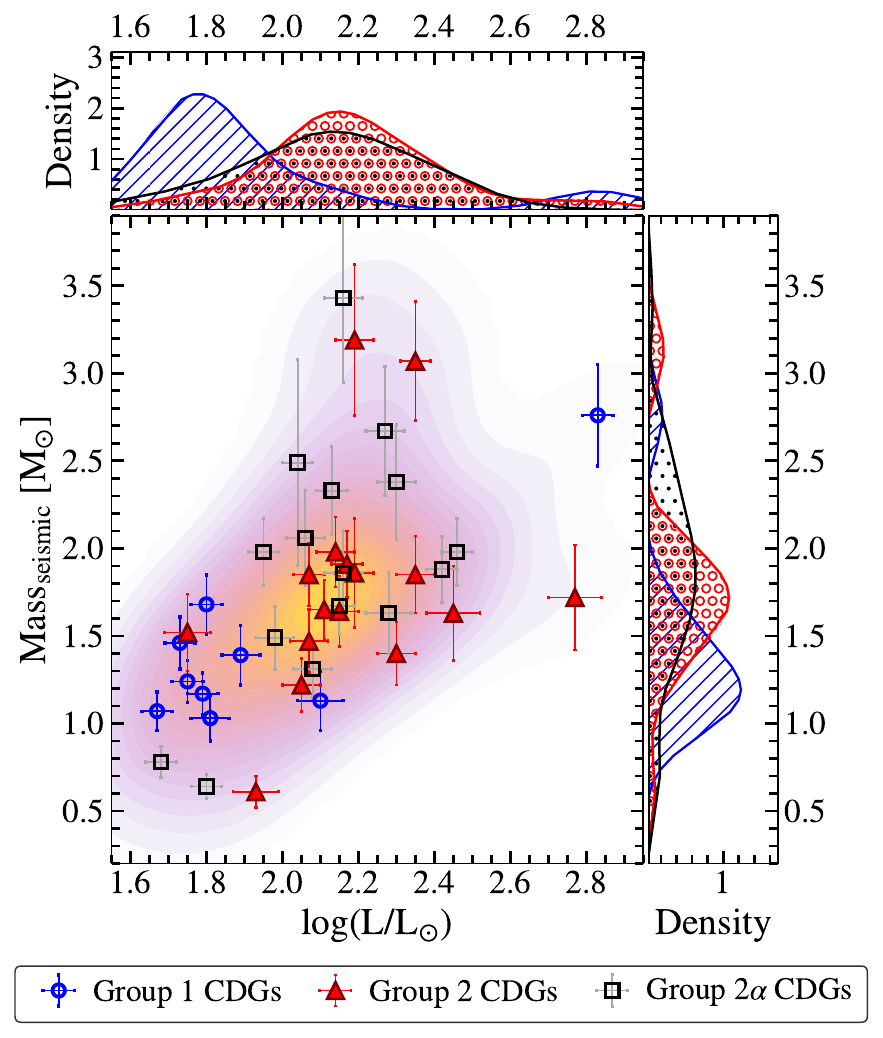}
    \caption{Seismic mass versus $\log \rm(L/L_{\odot})$ for CDGs, with Group~1, Group~2, and Group~2$\alpha$ shown as blue circles, red triangles, and black squares, respectively. Kernel density histograms illustrate the distributions for each group, using the corresponding group colors. Contours represent the density distribution of the full sample.}
    \label{fig:luminosity_vs_mass}
\end{figure}

\begin{figure*}
    \centering
    \includegraphics[scale=5,width=\linewidth]{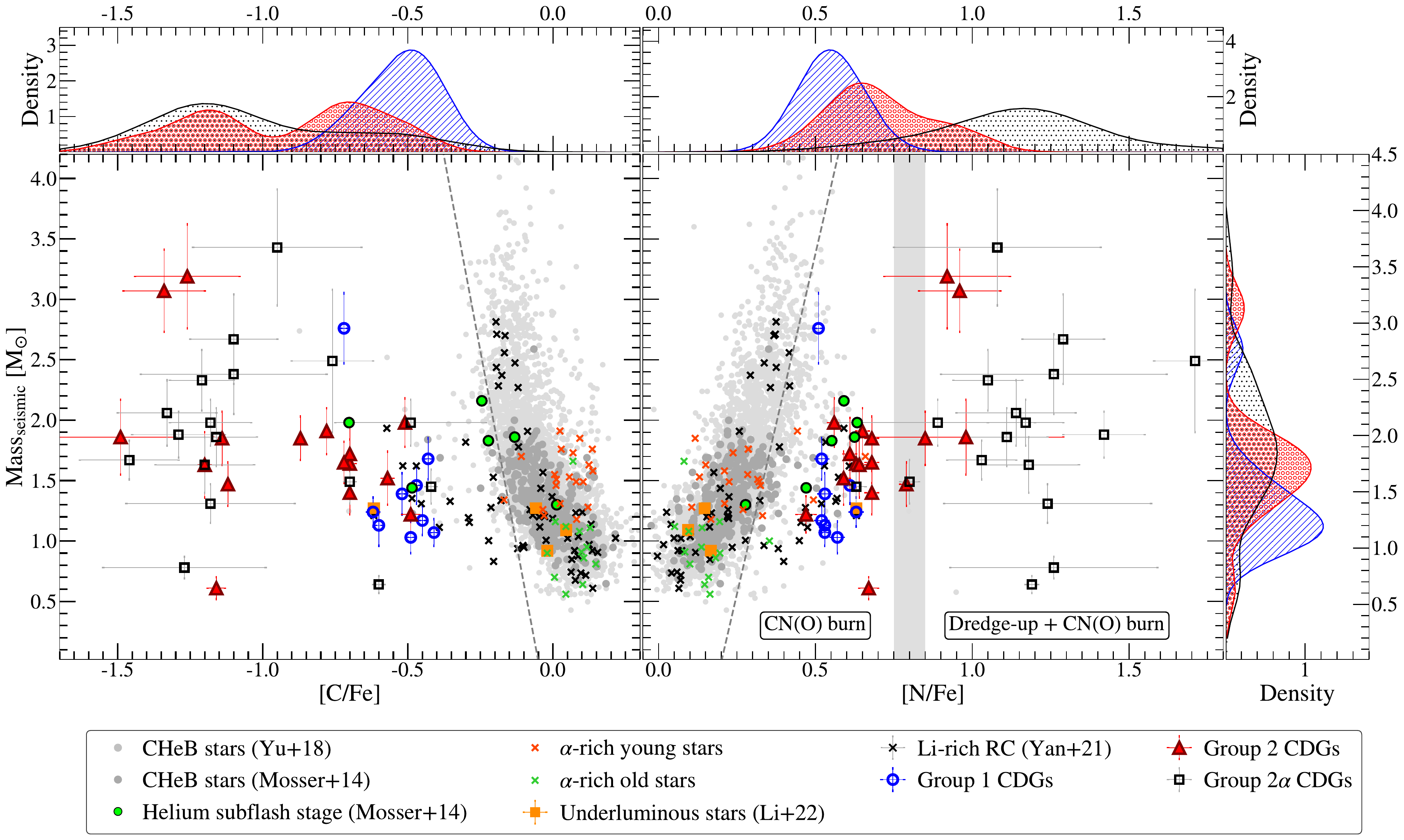}
    \caption{Seismic mass versus [C/Fe] (left panel) and versus [N/Fe] (right panel) for the CDGs, underluminous stars, Li-rich giants, $\alpha$-rich stars and a large sample of RC stars with symbols having the same meaning as in Figure~\ref{fig:sodium_vs_CNO}. We determine luminosities for the Li-rich giants in the same manner as our CDGs. Using this, we estimated the seismic masses of these stars using  mass Equation~\ref{eq:mass3}. The `normal' RC stars show a correlation between mass and nitrogen (and carbon) abundance. This reflects first dredge-up surface pollution from early on the RGB (e.g., \citealt{Iben1984}). The trend shows fairly sharp edges, allowing us to make a cut in N (or C) that varies with mass (dashed lines). This helps in distinguishing the chemically peculiar stars. To the right of the N line the stars can be considered N-rich, for their mass. The vertical shaded region at [N/Fe$] = +0.8\pm0.05$~dex, denotes the upper limit of our CDG sample’s N enhancement as well as the highest N abundance possible for scaled solar composition, if the ON cycle was activated after all C is burned to N.  Solar abundances are from \citet{Grevesse2007}.}
    \label{fig:mass_vs_carbon_nitrogen}
    \end{figure*}

On the other hand, comparing Group 2 and Group 2$\alpha$, we see that their surface temperatures and seismic $\log g$ are very similar. As mentioned above, luminosity is not a good discriminator between the groups, since there is overlap in luminosity distributions (see Figure~\ref{fig:luminosity_vs_mass}). Allowing for uncertainties, their masses, [Fe/H], [Na/Fe], $^{12}\rm{C}/^{13}\rm{C}$, and Li-rich fraction are also similar. Where Groups 2 and 2$\alpha$ diverge significantly is in their CNO abundances. Group 2$\alpha$ has, on average, much more extreme [C/Fe] depletion. This could be due to more thorough burning through the CN cycle. Although this suggests higher average [N/Fe] abundances -- which are observed -- the abundances are much higher than would be expected (+0.5~dex greater than the other groups). To illustrate this, in Figure~\ref{fig:mass_vs_carbon_nitrogen} we show an estimate for the highest [N/Fe] expected from a CN burn (vertical grey shading; assuming complete burning from initially scaled-solar composition). The majority of the Group 2$\alpha$ stars are found to have values greater than this upper limit, whilst the Group 2 CDGs are consistent with being at or below it. Looking at the \textit{distribution} of [C/Fe] in Figure~\ref{fig:mass_vs_carbon_nitrogen} we see that each distribution has (essentially the same) two peaks, and that there is a complete overlap between these two groups. The [C/Fe] averages are so different (Table~\ref{tab:CDG_groups_comparison}) because Group $2\alpha$ has more members in the peak around [C/Fe]~$\sim -1.25$~dex, than the peak around 0.7~dex. Apart from this the range and peak locations of the distributions are essentially the same. This begs the question as to why the Group $2\alpha$ CDGs have such high [N/Fe] when their [C/Fe] is similar to $\sim 40 \%$ of Group 2 CDGs. The similarity of the [C/Fe] distributions to Group 2 suggests that Group 2$\alpha$ has \textit{not} undergone more thorough burning through the CN cycle, reinforcing that N is particularly overabundant in these CDGs.

\begin{figure}
    \centering
    \includegraphics[width=1.0\linewidth]{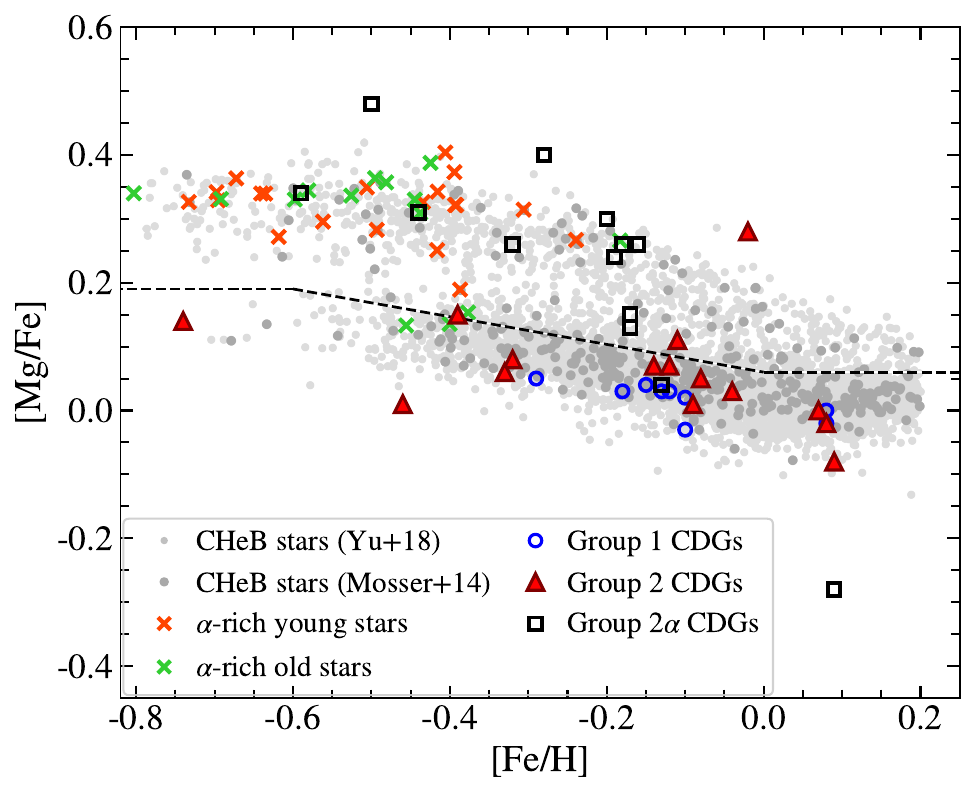}
    \caption{[Mg/Fe] ratio as a function of [Fe/H]. Giants classified based on asteroseismic analysis form the background from \cite{Mosser2014} and \cite{Yu2018} (small filled circles; see key). The $\alpha$-rich young (orange crosses) and old stars (green crosses) are compiled by \citet{Hekker2019}. Both the normal giants and these $\alpha$-rich stars have abundance ratios from APOGEE DR17. The Group~1, Group~2, and Group~2$\alpha$ CDGs are represented by blue circles, red triangles, and black squares, respectively. The black dashed line shows the separation between thin and thick disk stars based on the [Mg/Fe] ratio, similar to \cite{Adibekyan2011}.}
    \label{fig:mg_vs_metallicity}
\end{figure}

Turning to other elements, we see that Group $2\alpha$ has an overabundance of [O/Fe]. Overbundances of [O/Fe] between 0.2 and 0.5~dex are reminiscent of $\alpha$-enhanced stars. To check this we analysed other $\alpha$ element abundances for the sample. As an example, in Figure~\ref{fig:mg_vs_metallicity} we show [Mg/Fe] for all 3 groups, along with a background sample of Galactic giants, and a dedicated literature samples of $\alpha$-rich and $\alpha$-normal stars. From this figure it is clear that the Group $2\alpha$ stars can be associated with the $\alpha$-rich thick-disk population of the Galaxy (this is also seen in Ti), whilst the Group 2 (and Group 1) stars are predominantly $\alpha$-normal, following the thin disk composition. This strongly suggests that Group $2\alpha$ CDGs formed from a thick-disk-like initial composition, and that CN burning on this abundance pattern has led to the unusually high N. We checked this quantitatively by assuming that (i) no ON cycling had occurred (so the current [O/Fe] is the same as the initial), (ii) the initial [N/Fe] was scaled-solar (as seen in observations of $\alpha$-rich stars), and (iii) that the current [C+N+O/Fe] has not changed from the initial value. This gives an enhanced initial [C/Fe], which, when burned through the CN cycle, results in high N abundances, since C is much more abundant than N. Using this method on a star-by-star basis, we matched almost all of the [N/Fe] values currently observed in the Group $2\alpha$ CDGs. Since all other parameters are essentially the same as Group 2 (allowing for some uncertainties), we conclude that Group $2\alpha$ CDGs are just $\alpha$-rich counterparts of the Group 2 CDGs. That is, their origin is very likely to have been the same, they just formed with a different composition -- hence the name Group $2\alpha$. This interpretation is different to that in \cite{Maben2023b} -- where they concluded that some dredge-up from the core had enhanced the O -- because our current data set is much larger and more detailed. This finding reduces the complexity of the CDGs, suggesting that one less formation scenario is required.

\begin{table}[]
    \caption{Comparison of CDG classifications. Approximate modes are given for each parameter. Li-rich fractions are ranges, with lower bounds from measured Li-rich stars only and upper bounds assuming all unmeasured stars are Li-rich. Ellipses indicate unavailable data.}
    \label{tab:CDG_groups_comparison}
    \centering
    \begin{tabular}{@{}cccc@{}}
    \hline
     & Group~1 & Group~2 & Group~2$\alpha$ \\
        \hline
        \hline
        \multirow{1}{*}{$T_{\rm eff}$ (K)}
            & 4840 & 5000 & 5050 \\
        \addlinespace[0.01cm]
        \multirow{1}{*}{$\log g_{\text{seismic}}$}
         & 2.5 & 2.4 & 2.4 \\

        \addlinespace[0.01cm]
        \multirow{1}{*}{Lum. (L$_{\odot}$)}
            & 61 & 116 & 143 \\
        \addlinespace[0.01cm]
        \multirow{1}{*}{Mass (M$_{\odot}$)}
            & 1.2 & 1.7  & 1.9  \\

        \addlinespace[0.01cm]
        \multirow{1}{*}{[Na/Fe]}
            & 0.0 & $+0.3$  & $+0.4$ \\
        \addlinespace[0.01cm]
        \multirow{1}{*}{[CNO/Fe]}
            & $0.0$ & $0.0$ & $+0.3$ \\
        \addlinespace[0.01cm]
        \multirow{1}{*}{Li-rich frac.}
             & 67\%--100\% & 38\%--94\% & 53\%--59\%  \\

        \addlinespace[0.01cm]
        \multirow{1}{*}{[C/Fe]}
            & $-0.5$ & $-0.7$ & $-1.2$ \\
        \addlinespace[0.01cm]
        \multirow{1}{*}{[N/Fe]}
            & $+0.6$ & $+0.7$ & $+1.2$ \\
        \addlinespace[0.01cm]
        \multirow{1}{*}{[O/Fe]}
            & $0.0$ & $0.0$ & $+0.2$ \\
        \addlinespace[0.01cm]
        \multirow{1}{*}{[C/N]}
            & $-1.0$ & $-1.4$ & $-2.4$ \\
        \addlinespace[0.01cm]
        \multirow{1}{*}{$\rm^{12}C/^{13}C$}
            & ... & $5$ & $4$ \\
        \addlinespace[0.01cm]
        \multirow{1}{*}{[Fe/H]}
            & $-0.1$ & $-0.1$ & $-0.2$  \\
        \addlinespace[0.01cm]
    \hline
    \end{tabular}%
\end{table}

\subsubsection{Further analysis of chemical patterns of the three groups} \label{sec:further_analysis_of_chemical_pattern}

As noted above, Groups~1 and 2 have [C+N+O/Fe]~$= 0.0$, reflecting CN cycling without detectable enrichment \citep{Caughlan1965}. With our new analysis of Group~$2\alpha$ showing that those CDGs has different initial compositions, it is now clear that Group~$2\alpha$ also only underwent CN cycling, and no ON cycling. As pointed out by \cite{Maben2023b}, Group~1’s [Na/Fe]~$= 0.0$ indicates moderate-temperature burning ($20–40$~MK; \citealt{Adamczak2013}), and group~2’s [Na/Fe]~$= +0.3$, with [C/Fe]~$= -0.7$, [N/Fe]~$= +0.7$, [O/Fe]~$= 0.0$, suggests hotter conditions ($50–60$~MK). The sodium abundance in Group~$2\alpha$ is very similar, and, as discussed above, only CN cycling has occurred, suggesting these CDGs underwent burning at about the same temperature as Group 2.

We have $^{12}$C/$^{13}$C ratios for 16 of the 42 CDGs from the literature. The modes of the ratio distributions for Group 2 and Group $2\alpha$ are given in Table~\ref{tab:CDG_groups_comparison} (we have no data for Group 1 stars). Both values are low, 4-5, supporting advanced CN cycling, consistent with the [C/N] values \citep{Brown1989}.

   To further probe stellar processing, we analyse lithium abundances, for which we have measurements for 29 of the 42 CDGs (Table~\ref{tab:CDG_groups_comparison}). All 6 of 6 Group~1 stars with Li measurements are found to be Li-rich (A(Li)~$> 1.5$~dex), giving a Li-rich fraction of 67–100\% (3 stars have no measurements). For Group 2, 6 of 7 stars are Li-rich (38–94\%, 9 stars without measurements), and for Group $2\alpha$, 9 of 16 stars are Li-rich (53–59\%, 1 star with no measurement). These very high rates of Li-richness, despite high burning temperatures, points to mixing in the event(s) that formed the chemical patterns of the CDGs.

This prevalence of Li-rich giants is particularly remarkable given that, during the first dredge-up on the red giant branch, convective mixing dramatically reduces the surface lithium abundance---A(Li) typically drops by about 95\% from its main-sequence (MS) value \citep{Iben1967}.  As a result, Li-rich giants are extremely rare in the field, representing only about 1\% of giants \citep{Brown1989, Martell2013, Casey2019, Kumar2020}. The exceptionally high Li-rich fractions found in all three CDG groups thus stand in stark contrast to this general rarity, highlighting the unusual nature of these stars and the importance of internal mixing or non-standard processes in shaping their chemical compositions.

   Finally we note that pollution from AGB stars as a source of the abundance patterns of CDGs is unlikely given that [C+N+O] appears to be unchanged in all groups, while [C+N+O] increases with third dredge-up in thermally pulsing AGB (TP-AGB) stars (e.g., \citealt{Karakas2014}).

    \subsubsection{Kinematics and Galactic Component Membership} \label{sec:kinematics}

    Following \citet{Maben2023b}, we analyzed CDG kinematics using Astropy’s \texttt{Galactocentric} module \citep{Astropy2013, Astropy2018}, with distances \citep{Bailer-Jones2021}, $Gaia$~DR3 proper motions \citep{Gaia2016, Gaia2022}, and radial velocities from literature \citep{Adamczak2013, Palacios2016, Holanda2023, Maben2023a, Maben2023b, Holanda2024} and \citet{Luck2007} for the CDGs and local giants, respectively.  We categorize stars with $v_{tot} < 70$\,km\,s$^{-1}$ as thin disk stars, those between $70$ and $180$\,km\,s$^{-1}$ as thick disk stars, and those with velocities greater than this as halo stars \citep{Nissen2004, Venn2004}.

\begin{figure*}[]
    \centering
    \includegraphics[width=1.0\linewidth]{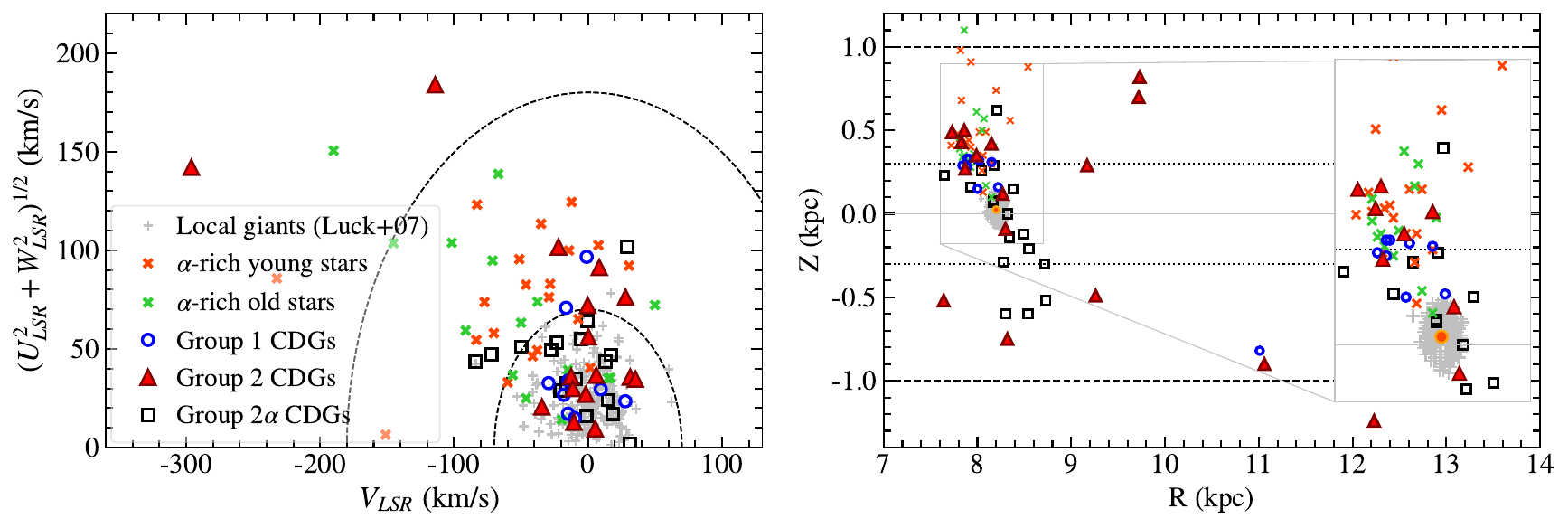}
    \caption{Left panel: The three groups of CDGs, including $\alpha$-rich young and old stars compiled by \citet{Hekker2019} and local giants from \citet{Luck2007}, are shown in the Toomre diagram. Dashed lines show constant values of the total space velocity, $v_{tot}$ = ($U^{2}_{LSR}$+$V^{2}_{LSR}$+$W^{2}_{LSR}$)$^{1/2}$ at 70 and 180\,km\,s$^{-1}$, demarcating thin disk and thick disk components, respectively \citep{Nissen2004, Venn2004}. Right panel: Spatial distribution above/below the Galactic plane vs. the galactocentric distance. The commonly accepted scale heights of the thin disk and thick disk are indicated at $Z = \pm 0.3$~kpc (dotted lines) and  $Z = \pm 1$~kpc (dashed lines), respectively \citep{Sparke2007}. The Sun (filled yellow circle) is at R = 8.2~kpc and Z = 0.025~kpc.}
    \label{fig:kinematics}
    \end{figure*}

    The Toomre diagram, shown in the left panel of Figure~\ref{fig:kinematics}, reveals that 78\% of Group~1 CDGs belong to the thin disk, with two CDGs classified as thick disk members.  For Group~2, 63\% of the CDGs are thin disk members, four CDGs are thick disk members, and two CDGs belong to the halo population. Group~2$\alpha$ CDGs are 76\% thin disk members, while four CDGs belong to the thick disk. Interestingly, old thin-disk stars with enhanced [$\alpha$/Fe] at similar [Fe/H] share kinematic properties with Group~2$\alpha$ CDGs \citep{Nepal2024}. Despite exhibiting high [$\alpha$/Fe] ratios typical of the thick disk, the kinematics of Group~2$\alpha$ CDGs indicate that they predominantly occupy thin disk orbits.

    Overall, our kinematic analysis shows that approximately 72\% of the CDGs belong to the thin disk, 23\% to the thick disk, and 5\% to the halo. The magnitude-limited nature of our sample favors nearby thin disk stars, likely influenced by dynamical mixing near the thin-thick disk boundary. The spatial distribution in the $R$--$Z$ plane (see right panel of Figure~\ref{fig:kinematics}) further illustrates the spread expected from dynamical evolution \citep{Sparke2007, Bond2019}.

    Although chemically peculiar, the kinematics of the CDGs provide valuable insights into the dynamical history of the Milky Way’s disk and its distinct stellar populations.

\subsection{Formation scenarios for CDGs}
\label{sec:formation_scenarios}

The formation mechanisms of CDGs remain a subject of active investigation. \cite{Maben2023b} identified three groups of CDGs based on their luminosity and chemical properties: (i) normal-luminosity RC CDGs, (ii) overluminous RC CDGs, and (iii) overluminous, highly polluted CDGs (previously known as wGb stars). They found that mergers between HeWDs and RGB stars were the most likely formation pathway for the two overluminous groups, while binary mass transfer from intermediate-mass AGB stars remained a viable scenario for the highly polluted subset. For the normal-luminosity CDGs, they could not distinguish between pollution from the core He-flash and lower-mass merger events with the available data.

   \begin{figure*}[]
    \centering
    \includegraphics[width=0.8\linewidth]{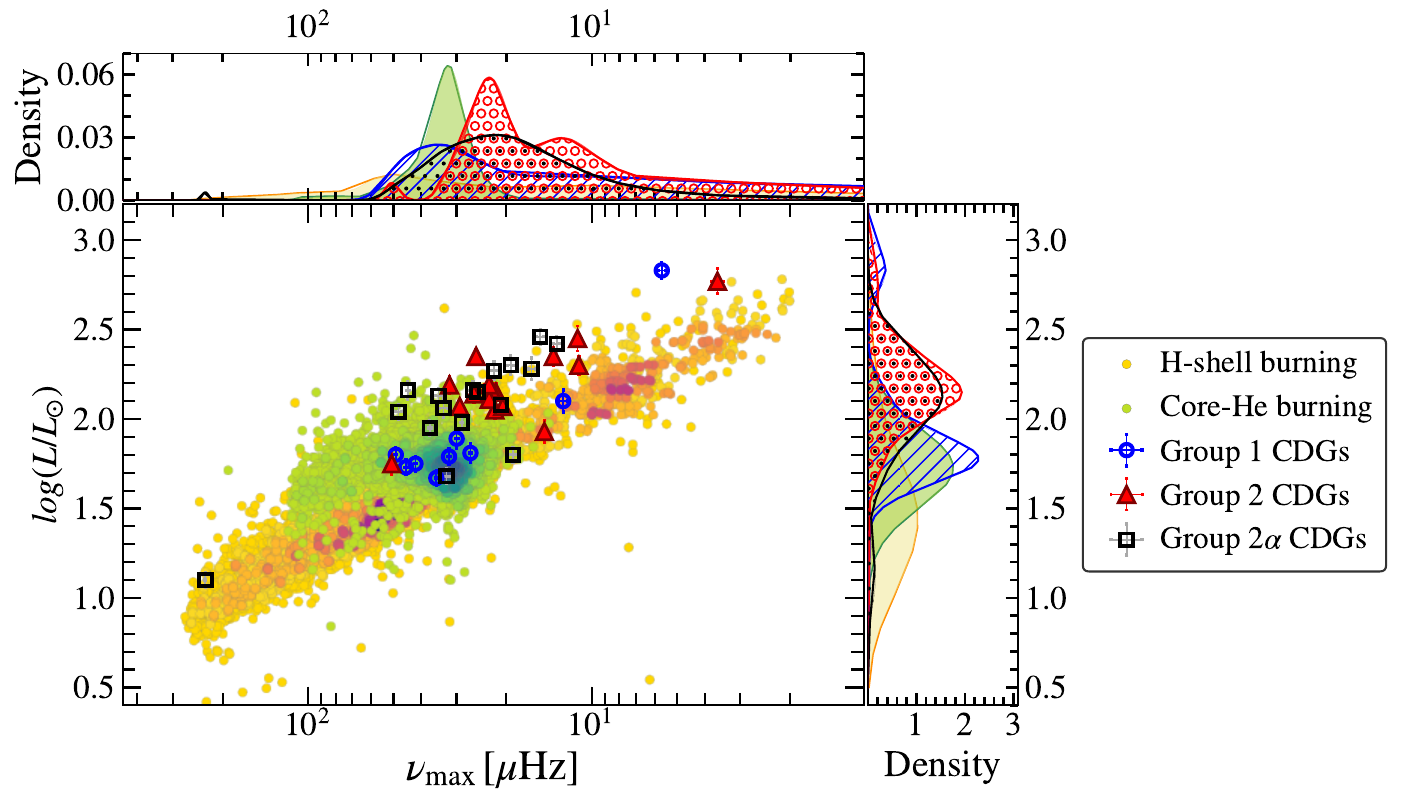}
    \caption{$\log(L/L_{\odot})$ is shown as a function of $\nu_{\mathrm{max}}$ for CDGs. Group~1, Group~2, and Group~2$\alpha$ stars are indicated by blue circles, red triangles, and black squares, respectively. RGB stars (yellow circles) and He-core burning giants (green circles) from \citet{Pinsonneault2018} and \citet{Yu2018} form the background, with the color scale representing the number density of stars (darker colors indicate higher density). Kernel density histograms for each group are shown along the axes, following the same color scheme. }
    \label{fig:luminosity_vs_numax}
    \end{figure*}

With the expanded CDG sample and improved asteroseismic constraints available in this study, we are now positioned to reassess the formation pathways previously proposed for the CDG groups. We evaluate the plausibility of each scenario in light of the updated seismic, photometric, and abundance information.

The $\log(L/L_\odot)$--$\nu_{\mathrm{max}}$ diagram (Figure~\ref{fig:luminosity_vs_numax}) serves as a useful diagnostic. When comparing the CDGs with the broader RC and RGB populations in this plane we see that the distribution of CDGs shows deviations from standard evolutionary sequences, offering insight into their formation histories.

\subsubsection{Group 1: Core-flash mixing or low-mass  mergers?} \label{sec:group1_origin}

Group~1 CDGs display luminosities and seismic parameters consistent with core He-burning RC stars. For the majority of these stars, both $\Delta\Pi_1$ and $\Delta\nu$ are measured. Specifically, 7 out of 9 Group~1 stars have both parameters available \citep{Mosser2014, Vrard2016,Maben2023b} -- firmly establishing their evolutionary status. In Figure~\ref{fig:luminosity_vs_numax}, Group~1 stars overlap with the region occupied by the general RC population in both luminosity and $\nu_{\mathrm{max}}$, further supporting their classification as low-mass core He-burning giants.

Chemically, Group~1 stars show moderate carbon depletion and nitrogen enhancement, with [C+N+O/Fe] remaining near solar. Their sodium abundances are approximately solar ([Na/Fe]~$\approx$~0.0), consistent with CN cycling at moderate temperatures ($\simeq 20-40$~MK) without activation of the Ne–Na cycle \citep{Adamczak2013}. This pattern suggests internal envelope mixing with moderate burning rather than deep, high-temperature nucleosynthesis.

A striking feature of Group~1 is the exceptionally high fraction of Li-rich stars: all six stars with lithium measurements are Li-rich (A(Li) $> 1.5$~dex), in stark contrast to the field giant population, where such stars are very rare (e.g., \citealt{Brown1989, Martell2013, Casey2019, Kumar2020}). Such enrichment is consistent with \textit{in-situ} lithium production near the RGB tip, as proposed for low-mass RC stars by \citet{Kumar2020}.

The luminosity distribution and mass range of Group~1 stars further constrain the nature of the mixing processes responsible for their chemical peculiarities. The fact that CDGs are generally not observed at luminosities below the RC (see Figures~\ref{fig:HRD_main} and \ref{fig:luminosity_vs_numax}) suggests that the pollution or mixing event responsible for their chemical peculiarities occurs near the RGB tip, as also noted by \citet{Maben2023b}. Given the mass range of Group~1 stars—8 out of 9 stars lie between 1.03~$M_\odot$ and 1.68~$M_\odot$ —these stars are expected to have experienced degenerate helium ignition in their cores, i.e., the core He-flash. While previous studies have suggested that envelope pollution during the core He-flash could increase surface carbon \citep[e.g.,][]{Deupree1987, Izzard2007, Mocak2009}, such carbon enhancement is incompatible with the carbon-deficient nature of Group~1 stars. Instead, the mild \textit{in-situ} mixing event at the core He-flash which was proposed to explain Li-rich giants (\citealt{Kumar2011, Kumar2020, Maben2023b}; also see modeling by \citealt{Schwab2020,Mori2021}), was  extended by \citet{Maben2023b} to include a regime of intermediate flash-induced mixing in which partial CN(O) burning enhances nitrogen and depletes carbon without affecting sodium or total [C+N+O] -- matching the Group~1 CDG chemical patterns. To date, such a mixing regime has not been fully explored in stellar models.

While internal mixing remains a plausible explanation, external pollution scenarios must also be considered. In particular, mass transfer from an AGB companion would increase [C+N+O/Fe], which is not observed (see Section~\ref{sec:further_analysis_of_chemical_pattern}), so AGB mass transfer is precluded. The Group~1 CDG chemical patterns could still have come from a different type of binary companion (although we struggle to find candidates). Binary information would help in exploring this possibility, however only one Group~1 star is currently identified as a binary, corresponding to a binary fraction of $\sim$11\%. However, this likely represents a lower limit due to observational incompleteness. At present, the available binary data are insufficient to conclusively favor or rule out any binary formation pathway.

The mass distribution of Group~1 is very similar to the general RC population, with the same peak at 1.2~$M_\odot$ (Table~\ref{tab:CDG_groups_comparison}). The absence of a bias to higher-mass stars, and the absence of a high-mass tail, disfavors a merger origin for this group. However, given the limited sample size, this conclusion remains tentative.

In summary, Group~1 CDGs are low-mass RC stars exhibiting lithium enrichment and mild CN processing without sodium enhancement. Their properties are best explained by \textit{in-situ} mixing associated with the core He-flash, consistent with theoretical scenarios proposed by \citet{Kumar2011, Kumar2020}, \citet{Schwab2020}, and \citet{Maben2023b}. The absence of carbon enhancement and [C+N+O] enrichment argues against AGB mass transfer, and the lack of a mass bias and high-mass tail makes mergers less likely. This contrasts with Group~2 CDGs, which are more massive, sodium-rich, and likely require a merger origin (Section~\ref{sec:group2_origin}). More complete binary surveys and larger samples will be essential to confirm the dominant formation mechanism for Group~1.


\subsubsection{Group 2 and 2$\alpha$: He subflashes, AGB mass transfer, or more massive mergers?} \label{sec:group2_origin}

As discussed in Section~\ref{sec:group2a}, we now consider the Group 3 stars of \cite{Maben2023b} as $\alpha$-enhanced analogues of Group~2. This simplifies the analysis of the CDGs since we are now essentially only dealing with two groups.
As reported in \cite{Maben2023b}, Group~2 CDGs are clearly more luminous and more massive than normal RC stars (also see Table~\ref{tab:CDG_groups_comparison}). They also show more advanced burning as compared to Group 1, with enhanced Na and lower [C/N]. Although [C/Fe] and [N/Fe] are different between Group 2 and 2$\alpha$, we determined in Section~\ref{sec:group2a} that this is solely due to their different initial compositions.

Only a few stars in each group have both $\Delta\Pi_1$ and $\Delta\nu$ measurements available from the literature \citep{Mosser2014, Vrard2016, Yu2018}, specifically 3 out of 16 Group~2 stars and 1 out of 17 Group~2$\alpha$ stars. For these stars, the seismic values are consistent with core He-burning RC evolution. Two of the Group~2 CDGs are classified as helium subflashing stars by \cite{Mosser2014}. Subflashes occur during a brief phase leading up the the main core He-flash. In the $\nu_{\mathrm{max}}$–$\log L$ plane, the majority of Group~2 and Group~2$\alpha$ stars form an offset sequence at higher luminosity (lower $\nu_{\mathrm{max}}$) than normal RC stars of the same seismic mass, matching the “over-luminous” RC component seen in previous studies.

A key new result from our expanded sample is that AGB pollution can now be clearly ruled out as a formation pathway for both Group~2 and 2$\alpha$. In all cases, [C+N+O] remains unchanged from the stars' initial values (see analysis of Group~2$\alpha$ in Section~\ref{sec:group2a}), while third dredge-up in TP-AGB stars increases the total CNO abundance through the mixing up of the intershell material which is rich in C (e.g., \citealt{Karakas2014}). There is also no evidence for an unusually high binary fraction in Group~2: only 2 out of 16 stars (12.5\%)  are reported as binaries in the literature. Due to data limitations this is only a lower limit, but if it were to hold for the whole group, it would further argue against an AGB mass-transfer origin. In contrast, Group~2$\alpha$ stands out with a quite a high binary fraction: 10 out of 17 stars (59\%) are reported as binaries in the literature. This suggests that binary interactions or mergers are especially important in the evolutionary history of Group~2$\alpha$ CDGs.

Of the 10 Group~2$\alpha$ CDGs identified as binaries, five have both angular separation ($\rho$) and distance ($d$) measurements available in the literature, allowing us to estimate their orbital separations (Table~\ref{tab:group2a_binary_separations_table}). The projected separation $a_p$ was calculated as:
\begin{equation}
    a_p = \rho \times d,
\end{equation}
\noindent where $\rho$ is in arcseconds and $d$ in parsecs, yielding $a_p$ in au \citep{Holberg2013}. To convert this projected quantity into a physical orbital separation, information on the orbital eccentricity and inclination is required--but these parameters are not available for our systems. As such, we adopt a statistical approach. \citet{Dupuy2011} use Monte Carlo simulations to compute distributions of correction factors that convert projected separations into semimajor axes, averaging over orbital orientations and eccentricities. From their Table~6, we adopt the median correction factor of 1.10, appropriate for systems with uniform eccentricity distributions and no discovery bias. The orbital separation is then estimated as:
\begin{equation}
    \widehat{a} = 1.10 \times a_p.
\end{equation}
\begin{table}[]
    \centering
    \caption{Estimated orbital separations for  Group~2$\alpha$ CDGs.}
    \begin{tabular}{lcccc}
        \hline
        Star & $\rho$ & $d$ & $a_p$ & $\widehat{a}$ \\
             & (arcsec) & (pc) & (au) & (au) \\
        \hline
        \hline
        HD~91805   & 17.80$^{a}$  & 251.23  & 4471.99  & 4919.19 \\
        HD~40402 & 12.40$^{a}$ & 597.89 & 7413.90 & 8155.29 \\
        HD~18636 & 20.78$^{b}$ & 361.45 & 7511.04 & 8262.14 \\
        HD~49960 & 12.70$^{a}$ & 592.43 & 7523.89 & 8276.28 \\
        HD~124721 & 47.23$^{b}$ & 789.54 & 37290.19 & 41019.21 \\
        \hline
    \end{tabular}
    \label{tab:group2a_binary_separations_table}
    {\small
    \vspace{0.5em}
    \textbf{Note:}
    $^{a}$ \cite{Mason2001}; $^{b}$ \cite{Badry2021}
    }
\end{table}

   \begin{figure}[]
    \centering
    \includegraphics[width=\linewidth]{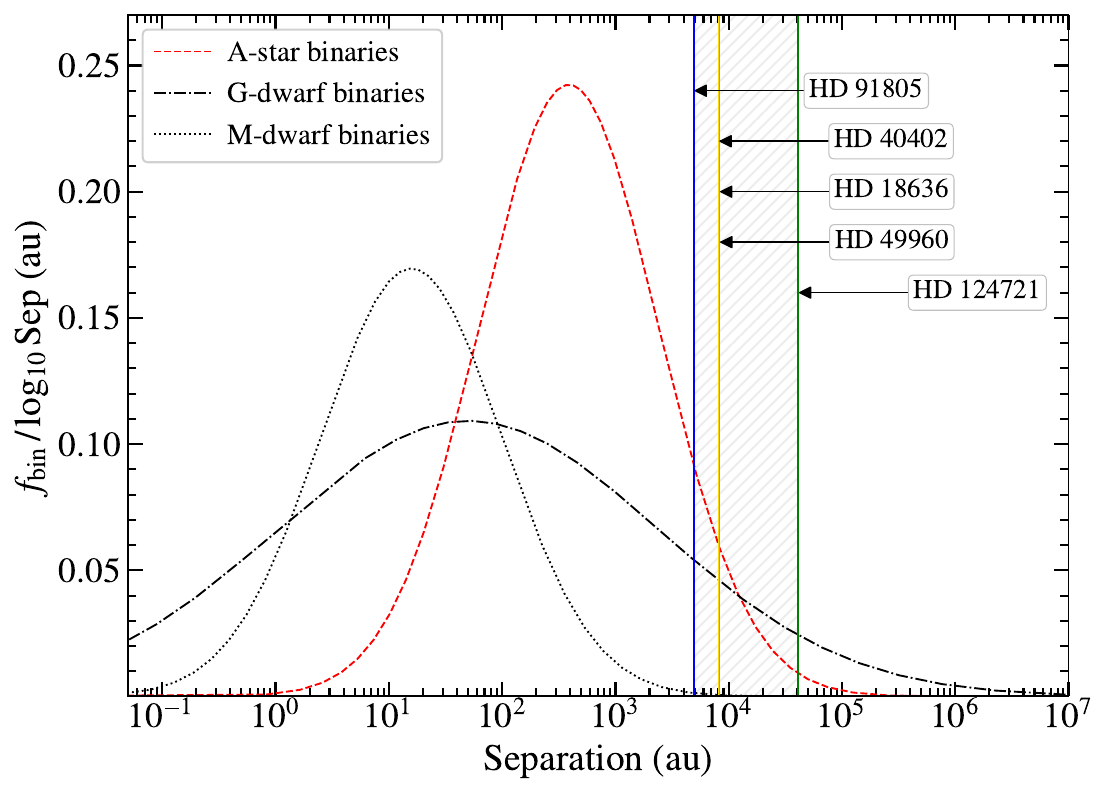}
    \caption{Log-normal fits to binary separation distributions in the Galactic field, following \citet{Parker2014}. The fit to A-star binaries (red dashed) is from \citet{DeRosa2014}, G-dwarfs (black dash-dotted) from \citet{Raghavan2010}, and  M-dwarfs (black dotted) from \citet{Janson2012}. Each distribution is normalized to the observed binary fraction for that stellar type. Vertical solid lines mark estimated orbital separations for 5 of the 10 Group~2$\alpha$ CDG binaries with available astrometric parameters (see text for details). The hatched region spans the range of these systems (4900--41000~AU), illustrating that they lie in the very wide binary regime.}
    \label{fig:group2a_binary_separation}
    \end{figure}

All five stars with available orbital data are found to be very wide binaries ($\sim 5000-40,000$~au; Table~\ref{tab:group2a_binary_separations_table}).

To examine whether CDGs have different orbital properties to normal field stars, we recreated the binary separation distributions from \citet{Parker2014} (see their Figure~1) and placed the estimated orbital separations of our Group~2$\alpha$ CDG binaries on the plot (Figure~\ref{fig:group2a_binary_separation}). The CDG binaries appear mostly at the (very) wide end of the separation range. This might suggest that these stars prefer very wide binaries. However, the current data are likely biased toward detecting wide binaries, so we cannot draw firm conclusions. Based on available data, 5/17 (29\%) of Group~2$\alpha$ CDGs are very wide binaries, which gives us a lower limit on the wide binary fraction. This could still be consistent with normal binary populations, but more data are needed to be sure.

The overall binary fraction should also be considered. Our sample shows a 59\% binary fraction for Group~2$\alpha$ CDGs. If this includes both close and wide binaries, the value may be directly comparable to the overall binary fraction in the field. But if the detections are mostly sensitive to wide binaries (as with the five systems above), the 59\% should be compared to the very wide binary fraction, in which case it is very high compared to that observed in field stars (e.g.,~\citealt{DeRosa2014,Parker2014}). That is, the Group~2$\alpha$ binary fraction would appear to be elevated. Only a dedicated binarity study can definitively tell us the true separation distribution, so this is an important topic for future work.

These wide CDG binaries can be explained if these stars arise from hierarchical triple systems. In this case, the inner two stars merge to form the CDG, while the third star remains in a wide orbit, with the system then identified as a wide binary. This idea is supported by the work of \citet{Shariat2025}, who used simulations including stellar evolution and orbital dynamics. Their results show that over half of solar-type triples merge the inner binary within 12.5~Gyr, often due to eccentric Kozai–Lidov oscillations. The merged star remains in a wide binary with the third star.

The \citet{Shariat2025} models match many features of Group~2$\alpha$ CDGs, including wide orbital separations, merger timescales from hundreds of millions to several billion years, and merger products that lie on the giant branch. These post-merger stars are said to have distinct asteroseismic, photometric, and even astrometric signatures. Many of the WD+MS merger products in their simulations are described as modified RG stars--with a dense core and an inflated envelope generated from the MS component--which may appear brighter than expected and show chemical signatures of internal processing. While detailed abundance predictions are not provided, the mixing of processed material from evolved components (e.g., WDs or RGB stars) naturally allows for enhanced N and Na, depleted C, and low $^{12}$C/$^{13}$C ratios—consistent with the observed properties of CDGs. Simulated merger channels include both WD+RG and WD+MS systems, which are capable of producing core He-burning stars with elevated luminosities and evolved surface compositions. As mentioned in \cite{Maben2023b}, detailed merger models using stellar evolution codes including nucleosynthesis can show chemical signatures similar to CDGs \citep[e.g.,][]{Zhang2013, Zhang2020}.

As identified in \cite{Maben2023b}, the mass distribution of Group~2 and 2$\alpha$ stars is a strong piece of evidence for these bright CDGs being merger products. They have an average mass around 1.6~$M_\odot$, with no stars below 1.0~$M_\odot$, and are more massive than typical red clump stars (Table~\ref{tab:CDG_groups_comparison} and Fig.~\ref{fig:luminosity_vs_mass}). One natural interpretation for this bias towards higher masses is that the extra mass comes from combining two lower-mass stars through a merger.

Finally, the kinematics of Group~2$\alpha$ stars show that they are $\alpha$-enhanced yet have thin-disk orbital motions. This combination, while unusual, is seen in other studies \citep[e.g.,][]{Nepal2024}, and may be explained by radial migration. Our magnitude-limited sample also contributes, as nearby stars often reflect overlapping disk populations due to dynamical mixing.

In summary, the available evidence strongly favors a formation scenario in which Group~2 and 2$\alpha$ CDGs arise predominantly from binary mergers occurring within hierarchical triple systems. The relatively high overall binary fraction of about 60\% supports this interpretation, since a pure binary merger population would be expected to have a near-zero binary fraction, as the binary would have merged, leaving only a single star. In the heirarchical triple scenario, the inner binary can merge but the original tertiary companion remains, forming  a wide binary system, as observed for at least 29\% of our Group $2\alpha$ sample. Alternative formation channels, such as AGB mass transfer or helium flash mixing, do not fully explain the observed chemical and orbital properties.

Given the incompleteness and observational biases in the current binary data, these conclusions remain tentative. Nonetheless, the hierarchical triple merger scenario currently provides the most consistent explanation for the properties of Group~2 and 2$\alpha$ CDGs. Future dedicated observations to better characterize binary fractions and orbital parameters will be essential to confirm or rule out this picture.
\section{Conclusion}

We have presented the first systematic asteroseismic analysis of the entire known population of CDGs, combining $Kepler$, K2, and TESS photometry for 129 stars. By measuring $\nu_{\rm max}$ and applying seismic scaling relations, we derived stellar masses and placed strong constraints on the evolutionary states of these chemically peculiar giants.

Our main findings are as follows:
\begin{enumerate}
    \item We detect solar-like oscillations in 43 CDGs and, by measuring $\nu_{\rm max}$ and applying seismic scaling relations, determine precise masses for these stars. The majority ($\sim$79\%) of these oscillating CDGs are low-mass ($M \lesssim 2~M_\odot$), in contrast to earlier suggestions that CDGs are predominantly intermediate-mass stars.

    \item Our asteroseismic analysis reveals that most CDGs have $\nu_{\rm max}$ values lower than typical RC stars, and their luminosity distribution is distinctly bimodal. Together with their temperatures, these results indicate that the majority of CDGs are in the core He-burning or EAGB phases, with only a single clear RGB candidate in our sample.

   \item Our expanded asteroseismic sample confirms the three chemically distinct groups of CDGs identified in \citet{Maben2023b}, most clearly distinguished in the [Na/Fe] versus [C+N+O/Fe] abundance plane: Group~1 (normal [Na/Fe], scaled-solar [C+N+O/Fe]), Group~2 (enhanced [Na/Fe], scaled-solar [C+N+O/Fe]), and Group~$2\alpha$ (enhanced in both). We find that Group~$2\alpha$ are the $\alpha$-enhanced analogs of Group~2, with enhanced initial C abundances explaining their extreme N enhancements.

 \item Lithium enrichment is prevalent across all groups, linking CDGs to the broader population of lithium-rich giants and suggesting a common evolutionary channel.

 \item We confirm that spectroscopic $\log g$ values are systematically offset from seismic values in CDGs, highlighting the necessity of asteroseismic constraints for accurate surface gravity and mass determination in these stars.

 \item For single CDGs, expected and observed $\nu_{\rm max}$ values agree well. In contrast, binaries and multiple systems show systematic discrepancies, likely due to contamination affecting spectroscopic parameters. This underscores the importance of accounting for binarity in asteroseismic and spectroscopic analyses.

\item Significant observational biases—especially due to brightness and frequency resolution in TESS data—limit the detectability of $\nu_{\rm max}$ in our sample.

\item Our analysis supports distinct formation pathways for the different CDG groups, based on their updated classifications and orbital properties:
    \begin{itemize}
        \item Group~1: Low-mass, Li-rich red clump stars consistent with internal mixing during the core He-flash. Their moderate CN processing and solar [Na/Fe] abundances suggest \textit{in situ} processes without external mass transfer. Their normal mass and luminosity distributions suggest they are not merger products.

        \item Groups~2 and 2$\alpha$: With a mass bias towards higher masses, these stars show strong chemical processing, with enhanced [Na/Fe] and strongly depleted [C/Fe]. Among the (identified) binaries in Group~2$\alpha$, a high fraction have very wide orbital separations. Together with the mass bias, this is consistent with formation through mergers in hierarchical triple systems, where the CDG originates from an inner binary merger and the tertiary companion remains gravitationally bound.

    \end{itemize}

\end{enumerate}

This work provides the first large-scale asteroseismic mass constraints and a refined evolutionary framework for CDGs, resolving long-standing ambiguities about their nature and formation. In future work, we will use measurements of the large frequency separation ($\Delta\nu$) and period spacing of mixed modes ($\Delta P$) to resolve remaining degeneracies in their evolutionary states and further constrain their internal structure. We are also constructing stellar merger models, which we will compare with the now substantial observational constraints. Continued time-domain photometry, high-resolution spectroscopy, and radial velocity monitoring will be important for fully characterizing the diversity and origins of this rare stellar population.


\section*{Acknowledgments}

This study is supported by the National Natural Science Foundation of China (NSFC) under grant No.12588202 and National Key R\&D Program of China No.2023YFE0107800, No.2024YFA1611900. SWC acknowledges federal funding from the Australian Research Council through a Future Fellowship (FT160100046) and Discovery Projects (DP190102431 and DP210101299). This research was partly supported by use of the Nectar Research Cloud, a collaborative Australian research platform supported by the National Collaborative Research Infrastructure Strategy (NCRIS). TRB acknowledges support from the Australian Research Council through Laureate Fellowship FL220100117. SM thanks Raghubar Singh and Zhuohan Li for helpful discussions. We thank Evgenii Neumerzhitckii for the use of his plotting routine.

This paper includes data collected by the $Kepler$ mission and obtained from the MAST data archive at the Space Telescope Science Institute (STScI). Funding for the $Kepler$ mission is provided by the NASA Science Mission Directorate. STScI is operated by the Association of Universities for Research in Astronomy, Inc., under NASA contract NAS~$5–26555$. This paper includes data collected with the TESS mission, obtained from the MAST data archive at the Space Telescope Science Institute (STScI). Funding for the TESS mission is provided by the NASA Explorer Program. STScI is operated by the Association of Universities for Research in Astronomy, Inc., under NASA contract NAS~$5–26555$.

This work has made use of data from the European Space Agency (ESA) mission $Gaia$ (\url{https://www.cosmos.esa.int/gaia}), processed by the $Gaia$ Data Processing and Analysis Consortium (DPAC, \url{https://www.cosmos.esa.int/web/gaia/dpac/consortium}). Funding for the DPAC has been provided by national institutions, in particular the institutions participating in the $Gaia$ Multilateral Agreement.

\software{\texttt{Astropy} \citep{Astropy2013, Astropy2018},
\texttt{Lightkurve} \citep{Lightkurve2018},
\texttt{Matplotlib} \citep{Matplotlib2007},
\texttt{NumPy} \citep{NumPy2020},
\texttt{Pandas} \citep{pandas2010},
\texttt{Python 3.12.2} \citep{python2009},
\texttt{SciPy} \citep{SciPy2020},
\texttt{TOPCAT} \citep{TOPCAT2005}
}


\appendix
\section{Asteroseismic parameter comparison}
\renewcommand{\thetable}{A\arabic{table}}
\setcounter{table}{0}

Table~\ref{tab:validation} lists literature asteroseismic parameters ($\nu_{\text{max}}$, $\Delta\nu$, $\Delta\Pi_1$) alongside our \texttt{pyMON}-derived $\nu_{\text{max}}$ values for comparison, for our sample of 27 CDGs.

\begin{longtable}{ccccccc}
\caption{Validation of asteroseismic parameters of the CDGs.} \label{tab:validation} \\
\hline
\multicolumn{7}{c}{\textit{CDGs having asteroseismic parameters derived from $Kepler$ light curves in the literature}} \\
\hline
Star & KIC & K$_{\text{p}}$ & $\nu_{\text{max, lit}}$$^{c}$ & $\Delta\nu_{\text{lit}}$ & $\Delta\Pi_{1, \text{lit}}$ & $\nu_{\text{max, pyMON}}$ \\
     &     &       & ($\mu$Hz) & ($\mu$Hz) & (s) & ($\mu$Hz) \\
\hline
\endfirsthead

 \multicolumn{7}{l}{\textbf{{\bfseries \tablename\ \thetable{}}} – \textit{continued}} \\
\hline
Star & KIC & K$_{\text{p}}$ & $\nu_{\text{max, lit}}$$^{c}$ & $\Delta\nu_{\text{lit}}$ & $\Delta\Pi_{1, \text{lit}}$ & $\nu_{\text{max, pyMON}}$ \\
     &     &       & ($\mu$Hz) & ($\mu$Hz) & (s) & ($\mu$Hz) \\
\hline
\endhead

2M19581582+4055411 & 5737930 & 9.95 & 3.49$\pm$0.3$^{f}$ & 0.62$^{f}$ &  & 3.61$\pm$0.17  \\

2M19252454+4036484 & 5446927  & 11.76 & 21.88$\pm$0.4  & 2.89$^{a}$ & 313$^{a}$ & 21.97$\pm$0.10   \\

2M19055092+3745351 & 2423824  & 11.05 & 22.02$\pm$0.5  & 2.70$^{a}$ & 360$^{a}$ & 21.79$\pm$0.12   \\

2M19090355+4407005 & 8222189  & 9.84  & 23.02$\pm$1.3  & 2.69$^{c}$ &     & 22.68$\pm$0.36    \\

2M19211488+3959431 & 4830861  & 11.11 & 25.94$\pm$0.5  & 2.96$^{c}$ &           & 26.02$\pm$0.14    \\

2M19382715+3827580 & 3355015  & 11.90 & 26.98$\pm$0.8  & 3.39$^{a}$ & 292$^{a}$ & 26.79$\pm$0.13   \\

2M19442885+4354544 & 8110538  & 12.85 & 28.77$\pm$0.9  & 3.62$^{b}$ & 334$^{b}$ & 28.76$\pm$0.15  \\

2M19400612+3907470 & 4071012  & 10.94 & 29.20$\pm$0.5  & 3.03$^{b}$ & 276$^{b}$ & 29.21$\pm$0.14  \\

2M19340082+4108491 & 5881715  & 11.46 & 30.90$\pm$1.1  & 3.42$^{a}$ & 202$^{a}$ & 29.97$\pm$0.24  \\

2M19422093+5018436 & 11971123 & 11.08 & 32.47$\pm$0.8  & 3.88$^{b}$ & 324$^{b}$ & 31.87$\pm$0.24   \\

2M19404764+3942376 & 4667911  & 10.50 & 36.37$\pm$1.0  & 4.18$^{b}$ & 318$^{b}$ & 35.33$\pm$0.24  \\

2M19133911+4011046 & 5000307  & 11.23 & 42.16$\pm$0.6  & 4.74$^{a}$ & 324$^{a}$ & 41.85$\pm$0.18   \\

2M19181645+4506527 & 8879518  & 10.98 & 46.18$\pm$0.8  & 4.65$^{b}$ & 268$^{b}$ & 45.21$\pm$0.22    \\

2M19125144+3850261 & 3736289  & 10.85 & 49.94$\pm$1.3  & 4.99$^{a}$ & 307$^{a}$ & 49.09$\pm$0.24   \\

2M19565550+4330561 & 7848354  & 13.12 & 51.74$\pm$1.5  & 4.67$^{c}$ &     & 50.74$\pm$0.29   \\

2M19052312+4422242 & 8352953  & 11.68 & 230.27$\pm$2.9  & 16.73$^{c}$ &         & 230.13$\pm$0.40  \\
\hline
\multicolumn{7}{c}{\textit{CDGs having asteroseismic parameters derived from TESS light curves in the literature}} \\
\hline
Star & TIC & Tmag & $\nu_{\text{max, lit}}$ & $\Delta\nu_{\text{lit}}$ & N$_{\text{sectors}}$ & $\nu_{\text{max, pyMON}}$ \\
     &     &       & ($\mu$Hz) & ($\mu$Hz) &  & ($\mu$Hz) \\
\hline
BD+5 593 & 283623989 & 8.03 & 23.50$\pm$3.7$^{d}$ &  & 2 & 20.95$\pm$0.51  \\

HD 56438 & 134545196 & 7.06 & 23.80$\pm$3.7$^{d}$ &  & 7 & 23.03$\pm$0.86  \\

HD 18474 & 192247771 & 4.63 & 25.14$\pm$1.7$^{e}$ & 2.70$\pm$0.4$^{e}$ & 2 & 25.29$\pm$0.48   \\

2M05120630-5913438 & 358459098 & 9.91 & 25.40$\pm$3.0$^{d}$ &  & 9 &  18.95$\pm$0.42   \\

2M06022767-6209038 & 149989441 & 10.01 & 29.50$\pm$4.5$^{d}$ &  & 8 & 23.00$\pm$0.77  \\

HD 124721 & 242443733 & 8.60 & 29.90$\pm$2.9$^{d}$ &  & 2 & 34.72$\pm$1.03  \\

HD 201557 & 231638013 & 8.29 & 31.50$\pm$1.4$^{e}$ & 2.99$\pm$0.2$^{e}$ & 2 & 31.74$\pm$0.30  \\

HD 16424 & 441127020 & 8.62 & 31.60$\pm$2.3$^{d}$ &  & 2 & 37.27$\pm$0.34  \\

HD 18636 & 321087542 & 6.80 & 36.60$\pm$2.4$^{d}$ &  & 4 & 32.52$\pm$0.69  \\

HD 40402 & 153122373 & 7.75 & 38.10$\pm$4.5$^{d}$ &  & 2 & 44.56$\pm$0.34  \\

HD 166208 & 332626441 & 4.19 & 48.73$\pm$2.7$^{e}$ & 4.35$\pm$0.5$^{e}$ & 7 & 48.03$\pm$1.26 \\

\end{longtable}
\vspace{-2em}
\begin{center}
{\small
    \textbf{Note:} All stellar designations beginning with `2M' denote the APOGEE ID of carbon-deficient giants.\\
    $^{a}$ \cite{Mosser2014}; $^{b}$ \cite{Vrard2016};
    $^{c}$ \cite{Yu2018};
    $^{d}$ \cite{Hon2021};
    $^{e}$ \cite{Zhou2024};    $^{f}$ \cite{Yu2020}
}
\end{center}

\section{Marginal and Non-Detections}
\renewcommand{\thetable}{B\arabic{table}}
\setcounter{table}{0}

Tables~\ref{tab:CDGs_marginal_detections} and \ref{tab:CDGs_no_detectable_oscillations} present atmospheric parameters for CDGs with marginal and non-detections of solar-type oscillations, respectively.

\begin{longtable}[c]{lrcccc}
    \caption{Atmospheric parameters of the CDGs with marginal detections. } \label{tab:CDGs_marginal_detections} \\

    \hline
    \multicolumn{1}{c}{Star} & \multicolumn{1}{c}{TIC} & Tmag & $T_{\rm eff}$ & $\log g$ & $\log(L/L_{\odot})$ \\
     &     &  & (K) &  & \\
    \hline
    \hline
    \endfirsthead

    \hline
    Star & TIC & Tmag & $T_{\rm eff}$ & log(g) & $\log(L/L_{\odot})$ \\
     &     &  & (K) &  & \\
    \hline
    \hline
    \endhead

    \hline
    \endfoot
    \hline
        HD 120213 & 418654329 & 4.55 & 4577 & 1.95 & 2.80 \\
        HD 119256 & 453698438 & 6.23 & 4984 & 2.64 & 2.89\\
        HD 67728 & 299844351 & 6.53 & 4827 & 2.28 & 2.82\\
        HD 215974 & 44088493 & 7.58 &  &  & 2.15 \\
        HD 132776 & 157884863 & 7.77 & 4680 & 2.30 & 2.10 \\
        2M19104331+2657415 & 407250258 & 10.44 & 5007 & 2.56 & 2.08\\
        2M05262269+2913054 & 285638910 & 11.55 & 4603 & 1.85 & 2.10  \\
        2M06282805-3106253 & 29629380 & 12.11 & 5037 & 2.25 & 2.24  \\
        2M01001653+6017239 & 256105839 & 12.40 & 4542 & 1.94 & 2.41 \\
        \hline
\end{longtable}
\vspace{-2em}
\begin{center}
{\small
    \textbf{Note:} All stellar designations beginning with `2M' denote the APOGEE ID of carbon-deficient giants.
}
\end{center}


\begin{longtable}[c]{lrcccc}
    \caption{Atmospheric parameters of the CDGs with no detectable oscillations.} \label{tab:CDGs_no_detectable_oscillations} \\

    \hline
    \multicolumn{1}{c}{Star} & KIC/EPIC/TIC & K$_{\text{p}}$/Tmag & $T_{\rm eff}$ & $\log g$ & $\log(L/L_{\odot})$ \\
         &     &  & (K) &  & \\
    \hline
    \hline
    \endfirsthead

    \multicolumn{6}{l}{\textbf{{\bfseries \tablename\ \thetable{}}} – \textit{continued}} \\

    \hline

    \multicolumn{1}{c}{Star} & KIC/EPIC/TIC & K$_{\text{p}}$/Tmag & $T_{\rm eff}$ & $\log g$ & $\log(L/L_{\odot})$ \\
         &     &  & (K) &  & \\
    \hline
    \hline
    \endhead

    \hline
    \endfoot
    \hline
        2M19522952+4210399$^{*}$ & 6717417 & 12.10 & 5111 & 2.86 & 1.69\\
         \hline
            HD 120170    & 212678022 & 8.87 & 5127 & 2.76 & 2.07\\
         \hline
            37 Com & 165945199 & 3.82 & 4610 & 2.50 & 2.66  \\
HD 26575 & 168751697 & 5.50 & 4690 & 2.20 & 2.39  \\
HD 21018 & 302769337 & 5.63 & 5310 & 1.60 & 2.39  \\
HD 31274 & 161476030 & 6.24 &  &  & 2.34  \\
HD 20090 & 441510543 & 6.96 &  &  & 2.44  \\
HD 28932 & 452783241 & 7.00 & 4915 & 2.50 & 2.30  \\
HD 17232 & 36819972 & 7.14 &  &  & 3.07  \\
HD 36552 & 354518979 & 7.28 &  &  & 2.06  \\
CD-2875 & 246820962 & 7.43 &  &  & 3.35  \\
HD 30297 & 286451983 & 7.53 &  &  & 2.00  \\
HD 198718 & 91434822 & 7.76 & 4980 & 2.50 & 2.10  \\
HD 204046 & 211407433 & 7.90 & 4984 & 2.51 & 2.27  \\
HD 105783 & 22026413 & 8.00 & 4860 & 2.10 & 2.20  \\
HD 207774 & 883711 & 8.06 & 5125 & 2.79 & 2.09  \\
CD-377613 & 165534155 & 8.95 &  &  & 2.04  \\
2M04495090+3851461 & 187205131 & 9.36 & 4863 & 2.34 & 2.22  \\
BD -19 967 & 121564557 & 9.36 &  &  & 1.82  \\
2M07581022+4325398 & 141071795 & 9.40 & 4969 & 2.42 & 2.04  \\
2M02491557+5147498 & 428071324 & 9.57 & 4687 & 2.04 & 2.67  \\
2M08445468+2957159 & 117280852 & 9.71 & 5052 & 3.07 & 1.39  \\
2M06362656-6347360 & 293347406 & 10.00 & 5221 & 2.42 & 2.17  \\
2M03370386+6830247 & 275911107 & 10.27 & 5220 & 2.48 & 1.91  \\
2M23540193+5649215 & 444325050 & 10.53 & 4245 & 1.22 & 3.01  \\
2M16055395+2356168 & 236323154 & 10.83 & 4969 & 2.42 & 2.25  \\
2M23065670+4724235 & 173298845 & 11.00 & 4795 & 2.07 & 2.26  \\
2M05524283+2648348 & 77824724 & 11.20 & 3955 & 0.76 & 2.68  \\
2M05323558+3213532 & 115055396 & 11.33 & 4451 & 1.44 & 2.34  \\
2M05294163+3942524 & 368147062 & 11.41 & 4989 & 2.22 & 2.32  \\
2M05501847-0010369 & 176615527 & 11.47 & 4070 & 0.94 & 2.89  \\
2M20564714+5013372 & 313233284 & 11.53 & 4059 & 1.11 & 2.51  \\
2M07514685-6416532 & 410446741 & 11.54 & 4986 & 3.06 & 1.40  \\
2M04482937+6336236 & 66216687 & 11.55 & 4995 & 2.51 & 1.82  \\
2M06572928-2942235 & 62965851 & 11.56 & 4860 & 2.10 & 2.08  \\
2M07084597-6227180 & 349056525 & 11.62 & 4987 & 2.22 & 2.15  \\
2M05491338-6528536 & 149627133 & 11.74 & 4795 & 2.47 & 1.74  \\
2M04572716+2503506 & 60604623 & 11.82 & 5012 & 2.24 & 2.03  \\
2M00290504+4952279 & 202643399 & 11.93 & 4841 & 2.35 & 1.82  \\
2M07442293-2344132 & 140539531 & 11.94 & 5074 & 2.65 & 2.00  \\
2M23552546+7455247 & 426145258 & 12.05 & 4965 & 2.46 & 1.78  \\
2M03345798+6800254 & 85508309 & 12.12 & 4577 & 1.66 & 2.52  \\
2M00254993+7357163 & 407461210 & 12.13 & 5045 & 2.53 & 2.17  \\
2M00220008+6915238 & 407116170 & 12.17 & 4098 & 1.08 & 2.57  \\
2M04303772+6042173 & 65781276 & 12.29 & 4886 & 2.46 & 1.80  \\
2M06383118+4646336 & 190378881 & 12.34 & 4858 & 2.16 & 2.04  \\
2M01450782+6445416 & 399264249 & 12.38 & 4392 & 1.63 & 2.29  \\
2M21184119+4836167 & 63158860 & 12.48 & 4199 & 1.25 & 1.63  \\
2M02073982+3707297 & 184653921 & 12.52 & 4993 & 2.16 & 2.31  \\
2M04592354-5907517 & 220460842 & 12.58 & 4960 & 2.96 & 1.42  \\
2M05481354+2926231 & 76540894 & 12.64 & 4720 & 2.12 & 2.19  \\
2M03394392+6932532 & 276033754 & 12.68 & 5005 & 2.37 & 1.91  \\
2M04403830+2554274 & 150095560 & 12.75 & 4992 & 2.41 & 1.78  \\
2M05261122+2914086 & 285471851 & 12.76 & 4913 & 2.13 & 2.08  \\
2M02461527+6024099 & 50766398 & 12.81 & 5092 & 2.52 & 1.87  \\
2M06475248+1837026 & 57130767 & 12.86 & 4803 & 1.85 & 2.61  \\
2M07061526-1526100 & 148579421 & 12.97 & 4601 & 1.94 & 2.26  \\
2M19181713+2703269 & 404209613 & 12.97 & 4641 & 2.17 & 2.25  \\
2M12261419-6325497 & 450565123 & 12.98 & 4667 & 2.01 & 4.66  \\
2M21405264+6021011 & 406519141 & 13.01 & 5033 & 2.44 & 2.26  \\
2M12254845-6228451 & 450475155 & 13.06 & 4820 & 2.43 & 2.73  \\
2M23461558-1711228 & 2762370 & 13.14 & 4308 & 1.16 & 2.57  \\
2M03431471+6657575 & 276191898 & 13.18 & 5124 & 2.44 & 1.91  \\
2M05155141+4608365 & 368180782 & 13.26 & 4987 & 2.52 & 2.42  \\
2M04154135-7223426 & 33835137 & 13.38 & 5337 & 2.65 & 1.84  \\
2M06242370+0556213 & 234571169 & 13.49 & 5007 & 2.61 & 2.14  \\
2M10005815-5533520 & 462058180 & 13.69 & 4986 & 2.12 & 3.60  \\
2M07171291-3935510 & 22974233 & 13.81 & 4949 & 2.09 & 2.00  \\
2M06144688+1544582 & 434369521 & 13.87 & 4706 & 2.12 & 1.88  \\
2M08234846-4918149 & 388755202 & 14.10 & 5012 & 2.14 & 2.33  \\
2M10552602-5856257 & 459848103 & 14.12 & 4980 & 2.32 & 3.52  \\
2M23041065-2410260 & 12970956 & 14.24 & 4480 & 1.43 & 1.79  \\
2M10042811-5526077 & 462394241 & 14.35 & 4939 & 2.47 & 3.48  \\
2M06295110+1158423 & 207012381 & 14.68 & 4981 & 2.43 & 1.62  \\
2M06414471+0138421 & 301589301 & 14.82 & 5055 & 2.94 & 1.66  \\
2M10041347-5515019 & 462361790 & 15.26 & 4942 & 2.55 & 2.76  \\
2M09102427-4824543 & 355666181 & 15.44 & 4457 & 1.56 & 2.64  \\

\end{longtable}
\vspace{-1em}

{\small
    \textbf{Note:} All stellar designations beginning with `2M' denote the APOGEE ID of carbon-deficient giants.
     The first row contains $Kepler$ Input Catalog (KIC) ID, the second row contains the K2 Ecliptic Plane Input Catalog (EPIC) ID, while the remaining rows contain TESS Input Catalog (TIC) IDs. 
     Horizontal lines are used to visually separate stars observed with $Kepler$ and K2 from those observed with TESS.
    $^{*}$ Pulsating star.

}

\bibliography{bibliography}{}
\bibliographystyle{aasjournal}

\end{document}